\documentclass[prd,aps,preprint,showpacs,superscriptaddress]{revtex4}

\usepackage{amsmath,amssymb}  
\usepackage{bm}               
\usepackage{graphicx}         
\usepackage{epsfig}           

\begin{document}

\preprint{BNL-NT-07/25} \preprint{RBRC-676}

\title{Single Transverse-Spin Asymmetry in Hadronic Dijet Production}

\author{Jian-Wei Qiu}
\email{jwq@iastate.edu}
\affiliation{Department of Physics and Astronomy,
             Iowa State University, Ames, IA 50011}
\affiliation{Physics Department, Brookhaven National Laboratory,
             Upton, NY 11973}

\author{Werner Vogelsang}
\email{vogelsan@quark.phy.bnl.gov}
\affiliation{Physics Department, Brookhaven National Laboratory,
Upton, NY 11973}

\author{Feng Yuan}
\email{fyuan@quark.phy.bnl.gov}
\affiliation{RIKEN BNL Research Center, Building 510A, Brookhaven
National Laboratory, Upton, NY 11973}
\date{\today}

\begin{abstract}
We study the single transverse-spin asymmetry for dijet
production in hadronic collisions in both the collinear QCD
factorization approach and the Brodsky-Hwang-Schmidt model. We
show that a nonvanishing asymmetry is generated
by both initial-state and final-state interactions, and that the
final-state interactions dominate. We find that in the leading
kinematic region where the transverse momentum imbalance of the
two jets, $\vec{q}_\perp = \vec{P}_{1\perp}+\vec{P}_{2\perp}$, is
much less than the momentum of either jet, the contribution from
the lowest non-trivial perturbative order to both the spin-averaged
and the spin-dependent dijet cross sections can be factorized into a
hard part that is a function only of the averaged jet momentum
$\vec{P}_\perp = (\vec{P}_{1\perp}-\vec{P}_{2\perp})/2$, and
perturbatively generated transverse momentum dependent (TMD)
parton distributions.
We show that the spin asymmetry at
this non-trivial perturbative order can be described by the TMD
parton distributions defined in either semi-inclusive DIS or the
Drell-Yan process. We derive the same hard parts
from both the collinear factorization approach and in the
context of the Brodsky-Hwang-Schmidt model, verifying that they are
not sensitive to details of the factorized long distance physics.
\end{abstract}


\maketitle

\newcommand{\be}{\begin{equation}}
\newcommand{\ee}{\end{equation}}
\newcommand{\ben}{\[}
\newcommand{\een}{\]}
\newcommand{\beqn}{\begin{eqnarray}}
\newcommand{\eeqn}{\end{eqnarray}}
\newcommand{\Tr}{{\rm Tr} }

\section{Introduction}

Single-transverse spin asymmetries (SSAs) in high-energy collisions
with one transversely polarized hadron are important phenomena that
have been observed for more than three decades in various physical
processes \cite{Bunce,E704,hermes,dis,star,phenix,brahms}.
In these processes, the observed final-state hadrons show an asymmetric
distribution in a plane perpendicular to the beam direction when
the transversely polarized hadron scatters off an unpolarized hadron
(or a virtual photon).  The SSA is defined as
$A_N\equiv (\sigma(S_\perp)-\sigma(-S_\perp)) /
           (\sigma(S_\perp)+\sigma(-S_\perp))$,
the ratio of the difference and the sum of (differential) cross
sections when the hadron's spin vector, $S_\perp$, is flipped.
Recent experimental measurements of SSAs in polarized
semi-inclusive lepton-nucleon deep inelastic scattering (SIDIS)
\cite{hermes,dis} and in hadronic collisions
\cite{star,phenix,brahms} have renewed the theoretical interest
in SSAs and in understanding their roles in hadron
structure and Quantum Chromodynamics (QCD).  Although it was
realized a long time ago \cite{KPR} that perturbative QCD can be
used to study the effects of transverse spin, the size of the
observed asymmetries came as a surprise and has posed a challenge
for researchers in this field \cite{review}.

Within a model calculation \cite{BroHwaSch02}, Brodsky, Hwang, and Schmidt
showed that the final state interaction in deep inelastic
scattering (DIS) can generate a phase required for a nonzero SSA
in SIDIS.  It was later realized that this final state interaction
can be factorized into the gauge link of the gauge invariant
transverse momentum dependent (TMD) quark distributions
\cite{Col02}. The nonvanishing SSA obtained in
Ref.~\cite{BroHwaSch02} is a consequence of the existence of a
naively time-reversal-odd TMD quark distribution, the so-called
Sivers function \cite{Siv90}. When applying the same calculation
to the Drell-Yan production of lepton pairs at hadron colliders,
the final state interaction in SIDIS becomes an initial state
interaction in Drell-Yan, and the phase changes sign, which leads
to a prediction of a sign change in the SSAs between these two
processes \cite{BroHwaSch02,Col02}. This nontrivial
``universality'' property associated with the TMD parton
distributions is the consequence of gauge interactions in QCD
\cite{Col02,BelJiYua02,BoeMulPij03}, and of the QCD factorization
theorems for these two processes
\cite{ColSop81,ColSopSte85,JiMaYu04,ColMet04}. Experimental tests
of this prediction will be crucial for our understanding of the
origin of SSAs in QCD~\cite{rhic-dy}.

In Ref.~\cite{BoeVog03}, it was proposed to study the Sivers
functions by means of a SSA in azimuthal-angular correlations of two jets
produced nearly back-to-back at hadron colliders.
Measurements of this SSA for dijet production have begun at RHIC
\cite{star-dijet1}, complementing the measurements in SIDIS.
Unlike the SIDIS or Drell-Yan process, dijet production at
hadron colliders involves both initial and final state interactions
that may produce the phase needed for a nonvanishing SSA.
Consequently, the sign and the size of the asymmetry will depend
on the relative strength of these interactions.
Following the previous works on SIDIS and Drell-Yan, the authors of
Ref.~\cite{mulders} developed a systematic approach to
describe the role of initial- and final-state interactions in
generating SSAs in hadronic collisions, and they found that
summing all initial/final state interactions into the gauge link
of the TMD parton distributions leads to a very complicated
functional form of the gauge link. In particular, the TMD parton
distributions studied in dijet correlations in hadronic
scattering will have no connection to those in the SIDIS and
Drell-Yan processes, because their definitions are completely
different \cite{mulders}. One then has to question the
universality of the TMD parton distributions, and the predictive
power of perturbative QCD calculations, which relies on comparing
physical observables with the same factorized long-distance
physics while having different perturbatively calculable
short-distance dynamics.

In Ref.~\cite{qvy-short}, we briefly reported a new result for
the SSA in dijet production in hadronic collisions in
the twist-3 Efremov-Teryaev-Qiu-Sterman (ETQS) approach
\cite{et,qiusterman}. We considered the spin-dependent cross
section, $\Delta\sigma(S_\perp)=(\sigma(S_\perp)-\sigma(S_\perp))/2$,
for the process
\begin{equation}
A(P_A,S_\perp)+B(P_B)\rightarrow J_1(P_1)+J_2(P_2)+X\ ,
\end{equation}
with the jet momenta $P_1\equiv P+q/2$ and $P_2\equiv -P+q/2$.
When both $P_\perp$ and $q_\perp$, the transverse components of
the momenta $P$ and $q$, respectively, are much larger than
$\Lambda_{\rm QCD}$, a nonvanishing $\Delta\sigma(S_\perp)$ can
be generated by the ETQS mechanism in the collinear factorization
approach. In this framework, the SSAs are attributed to the
spin-dependent twist-three quark-gluon correlation functions, which
correspond to a quantum interference between different partonic
scattering amplitudes. Since the incoming partons are approximated
to be collinear to the corresponding incoming hadrons in this
approach, the momentum imbalance of the two jets is generated by
producing a three-parton final-state. In Ref.~\cite{qvy-short}, we
calculated the contribution from initial-state gluon radiation to
$\Delta\sigma(S_\perp)$ in the kinematic region where $P_\perp
\gg q_\perp \gg \Lambda_{\rm QCD}$. We presented the final result
for the leading contributions to $\Delta\sigma(S_\perp)$ in the
expansion of the partonic scattering in $q_\perp/P_\perp$
involving a hard $qq'\to qq'$ subprocess. In this paper, we
will provide the detailed derivations of this result, and we will
also present the full contributions from all other partonic
subprocesses at the same order.  We find that although both
initial-state and final-state interaction lead to a nonvanishing
SSA, the final-state interactions give the dominant contributions
to $\Delta\sigma(S_\perp)$.  We therefore expect that the SSA in
dijet production will have the same sign as the Sivers asymmetry
in SIDIS.

We find that at leading order in the $q_\perp/P_\perp$ expansion, the
perturbatively calculated partonic parts can be further factorized
into a single-scale ($P_\perp$) hard part and perturbatively
generated TMD parton distributions with transverse momenta
$k_\perp = {\cal O}(q_\perp)$. We also find that our
perturbatively calculated result is equal to the leading-order
term in the $\Lambda_{\rm QCD}/q_\perp$ expansion of the following
generalized TMD factorization formula \cite{qvy-short}:
\begin{eqnarray}
\frac{d\Delta\sigma(S_\perp)}
     {dy_1dy_2dP_\perp^2d^2\vec{q}_\perp}
&=&
\frac{\epsilon^{\alpha\beta}S_\perp^\alpha q_\perp^\beta}
     {\vec{q}^2_\perp}
\sum\limits_{ab}
\int d^2k_{1\perp}d^2k_{2\perp}d^2\lambda_\perp
\nonumber \\
&&\times
\frac{\vec{k}_{1\perp}\cdot \vec{q}_\perp}{M_P}\,
   x_a\, q_{Ta}^{\rm SIDIS}(x_a,k_{1\perp})\,
   x_b\, f_b^{\rm SIDIS}(x_b,k_{2\perp})
\label{e4}\\
&&\times
\left[S_{ab\to cd}(\lambda_\perp)\,
      H_{ab\to cd}^{\rm Sivers}(P_\perp^2)\right]_c\,
\delta^{(2)}(\vec{k}_{1\perp}+\vec{k}_{2\perp}+
\vec{\lambda}_\perp-\vec{q}_\perp) \, ,
\nonumber
\end{eqnarray}
where $\sum_{a,b}$ runs over all parton flavors, $H_{ab\to
cd}^{\rm Sivers}$ and $S_{ab\to cd}$ are partonic hard and soft
factors, respectively, and the $[\quad ]_c$ represents a trace in
color space between the hard and soft factors due to the color
flow into the jets \cite{Botts:1989kf,Kidonakis:1997gm}. The hard
factor in Eq.~(\ref{e4}) only depends on the single hard scale
$P_\perp$ in terms of partonic Mandelstam variables of the
reaction $ab\to cd$:
\begin{eqnarray}\hat{s}&=&(p_a+p_b)^2=x_ax_b s\
, \nonumber\\
\hat{t}&=&(p_a-p_c)^2=-P_\perp^2\left(e^{y_2-y_1}+1\right)\ ,
\nonumber\\
\hat{u}&=&(p_b-p_c)^2=-P_\perp^2\left(e^{y_1-y_2}+1\right)\
,\label{mandelstam}
\end{eqnarray}
 with
$x_a=\frac{P_\perp}{\sqrt{s}}\left(e^{y_1}+e^{y_2}\right)$,
$x_b=\frac{P_\perp}{\sqrt{s}}\left(e^{-y_1}+e^{-y_2}\right)$ and
$y_1$ and $y_2$ the rapidities of the two jets. In Eq.~(\ref{e4}),
$q_{T_a}^{\rm SIDIS}$ and $f_b^{\rm SIDIS}$ denote the
transverse-spin dependent TMD quark distributions (known as the
Sivers function) and the unpolarized TMD quark distribution,
respectively; these TMD parton distributions were chosen to
follow their definitions in the semi-inclusive DIS process. For
example, for a polarized proton with momentum
$P=(P^+,0^-,0_\perp)$ with $P^\pm=1/\sqrt{2}\left(P^0\pm
P^3\right)$ and transverse spin vector $\vec{S}_\perp$, the
TMD distribution for quark flavor $a$ can be defined through the
decomposition of the following matrix element
\cite{JiQiuVogYua06},
\begin{eqnarray}
{\cal M}_a
&=& \int\frac{P^+d\xi^-}{\pi}\,\frac{d^2\xi_\perp}{(2\pi)^2}\,
e^{-ix\xi^-P^++i\xi_\perp\cdot k_\perp}
\langle PS|\overline{\mit \psi}_a(\xi){\cal L}_v^\dagger(\infty;\xi)
          {\cal L}_v(\infty;0){\mit \psi}_a(0)|PS\rangle
\nonumber \\
&=&\frac{1}{2}
\left[
 f_a^{\rm SIDIS}(x,k_\perp) \gamma_\mu P^\mu
+\frac{1}{M_P}
 q_{Ta}^{\rm SIDIS}(x,k_\perp)
 \epsilon_{\mu\nu\alpha\beta} \gamma^\mu P^\nu k^\alpha S^\beta
+ \dots \right] \ ,\label{e3}
\end{eqnarray}
where $M_P$ is a hadronic mass scale introduced to keep the TMD
distributions $f_a$ and $q_{Ta}$ at the same mass
dimension, and the gauge link ${\cal L}$ is defined in a covariant
gauge as ${\cal L}_{v}(\infty;\xi) = \exp\left(-ig\int^{\infty}_0
d\lambda \, v\cdot A(\lambda v +\xi)\right)$ with the path link
extended to $+\infty$.  $v$ is a vector conjugate to the momentum
vector $P$.  Since we will work in a covariant gauge
throughout this paper, the vector $v$ could be chosen to be a
light-cone vector with $v^2=0$ and $v\cdot P=1$. An off-light-cone
vector ($v^2\neq 0$) will have to be used when high order
corrections are taken into account~\cite{JiMaYu04}. If we work in
a singular gauge, like the light-cone gauge, an additional gauge
link at the spatial infinity ($\xi=+\infty$) will have to be
included in order to ensure the gauge invariance of the above definitions
\cite{BelJiYua02}. We have chosen TMD parton distributions defined
in SIDIS because of the dominance of final-state interactions.
Choosing TMD parton distributions defined according to the
Drell-Yan process would change the sign of the partonic hard
factors, but not affect the overall sign of the physical cross
section. We note that the various factors in Eq.~(\ref{e3}), apart
from being functions of transverse momentum, also depend on the
renormalization and factorization scales, and especially on the
gluon rapidity cut-off when higher-order corrections are taken
into account~\cite{{ColSop81},{ColSopSte85},JiMaYu04}. The latter
dependence is governed by the Collins-Soper evolution
equation~\cite{ColSop81}, which leads to a resummation of large
logarithms of the form $\alpha_s^n
\ln^{2n-1}\left[P_\perp/q_\perp\right]$ in the perturbative
series~\cite{ColSop81,ColSopSte85}.

In order to evaluate the SSA for dijet production, we also
calculate in this paper the contributions from initial-state gluon
radiation to the spin-averaged dijet cross section,
$\sigma=(\sigma(S_\perp)+\sigma(S_\perp))/2$, in the same kinematic
region where $P_\perp \gg q_\perp$. Using the collinear
factorization approach, we find that like the spin-dependent case,
the leading contribution in the $q_\perp/P_\perp$ expansion of the
perturbatively calculated partonic scatterings can be further factorized
into a hard part at a single-scale $P_\perp$ and perturbatively
generated unpolarized TMD parton distributions with transverse
momenta $k_\perp = {\cal O}(q_\perp)$. Our leading contribution in
the $q_\perp/P_\perp$ expansion is equal to the leading-order term of
the following TMD factorization formula \cite{VogYua05},
\begin{eqnarray}
\frac{d\sigma^{uu}}
     {dy_1dy_2dP_\perp^2d^2\vec{q}_\perp}
&=&
\sum\limits_{ab}
\int d^2k_{1\perp}d^2k_{2\perp}d^2\lambda_\perp \,
x_af_a^{\rm SIDIS}(x_a,k_{a\perp})\,
x_bf_b^{\rm SIDIS}(x_b,k_{b\perp})
\nonumber\\
&& 
\times
\left[S_{ab\to cd}(\lambda_\perp)
      H_{ab\to cd}^{uu}(P_\perp^2)\right]_c \,
\delta^{(2)}(\vec{k}_{a\perp}+\vec{k}_{b\perp}
            +\lambda_\perp-\vec{q}_\perp) \; ,
\label{e2}
\end{eqnarray}
where the superscript ``$uu$'' indicates the scattering of an
unpolarized beam off an unpolarized target/beam. For consistency,
we again express the factorization formula in terms of
TMD parton distributions defined in the SIDIS process.
Actually, the unpolarized TMD parton distributions defined in SIDIS and
the Drell-Yan process are identical, because the unpolarized parton
distributions are invariant under the naive-time-reversal
transformation.

A key feature of any QCD factorization is that the perturbatively
calculated short-distance hard factors should not be sensitive to
details of the {\it factorized} long distance physics. In this
paper, we will derive all short-distance hard factors in
Eqs.~(\ref{e4}),(\ref{e2}) also by using the Brodsky-Hwang-Schmidt model
for SSAs \cite{BroHwaSch02}. We find that the hard-scattering
factors derived in this model are in fact the same as those
derived in the collinear factorization approach, despite of the
clear difference in the treatment of the nucleon. Both the
spin-averaged and the spin-dependent cross sections calculated
in this model are therefore consistent with the generalized factorization
formulas in Eqs.~(\ref{e4}),(\ref{e2}).

To fully investigate the above factorization formalism, we would
have to consider the contributions from gluon interactions and
radiations in all possible regions of the phase space, and to all
orders in perturbation theory, which is not what we are trying to
do in this paper.  Instead, as a step to test the above
factorization formalism, we will study the contribution at the
first non-trivial order with a single gluon radiated nearly
parallel to one of the incident nucleons.  From this study, we
will be able to show how to factorize this gluon into the TMD
parton distributions of the incoming nucleons, and to verify the
definition of the TMD parton distributions used in the generalized
factorization formulas. Our first order calculation at $P_\perp\gg
q_\perp\gg\Lambda_{\rm QCD}$ clearly shows the factorization of a
hard part at ${\cal O}(P_\perp)$ and that the radiation of a
collinear gluon can be absorbed into the relevant TMD parton
distributions. However, our work does not address the
factorization between the various TMD parton distributions at the
scale ${\cal O}(q_\perp)$. Our calculation should be regarded
merely as a ``necessary condition'' for such a factorization to
hold. Since the predictive power of perturbative-QCD calculations
involving measured hadrons relies on factorization, a full proof
or disproof of this factorization remains an important challenge
\cite{collinsqiu}.

The rest of this paper is organized as follows. In Sec.~II, we
calculate the partonic hard factors in a simple model-inspired
approach.  Using the quark-diquark model for the proton and the
Brodsky-Hwang-Schmidt model for SSAs, we extract the hard factors
by the factorization of the TMD parton distributions. We find that
the hard factors for the unpolarized scattering in Eq.~(\ref{e4})
are identical to the partonic differential cross sections $d\hat
\sigma/d\hat t$, and the ones for the transverse-spin dependent
cross section are different due to the initial- and final-state
color interactions. In Sec.~III, we calculate the SSA in the
collinear factorization approach, and show that the collinear
gluons parallel to the polarized proton can be factorized into the
Sivers function, while those parallel to the unpolarized proton
can be factorized into the unpolarized TMD parton distribution. We
also demonstrate that the hard factors calculated in both
approaches are the same and independent of the details of the
factorized long-distance physics. In Sec. IV, we extend our formulas
to the $q_\perp$-weighted SSA, and compare with previous results
in the literature. Finally, we summarize our paper in Section V.

\section{Calculation of the Hard Factors}

In this section, we will calculate the hard factors at ${\cal
O}(P_\perp)$ in the generalized factorization formulas in
Eqs.~(\ref{e4}),(\ref{e2}), based on a model in which an energetic
parton scatters off a nucleon that is made of a quark and a scalar
diquark \cite{BroHwaSch02}. We will follow the approach of Brodsky
et al.~\cite{BroHwaSch02} and incorporate the correct color
interactions between the gauge boson and the scalar diquark in the
proton wave function. We extract the hard factor by factorizing
the model-sensitive long-distance physics from the differential
cross section into the parton distributions. Because the
model-dependence was factorized into the parton distributions, the
extracted hard factors resulting from our calculations do not depend
on how the nucleon couples to the partons, and therefore, should be
model-independent.

\begin{figure}[t]
\begin{center}
\includegraphics[width=9cm]{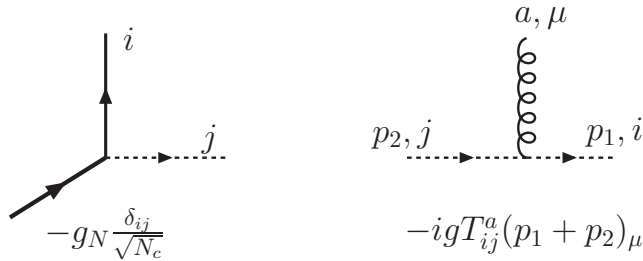}
\end{center}
\vskip -0.4cm \caption{\it  The vertices in the quark-diquark model
that we will use to calculate the hard factors. Here ``$a$" represents
the color-index for the gauge boson, and $i,j$ the color-indices
for the quark and/or the di-quark. }\label{f2}
\end{figure}

\subsection{Hard factors for the unpolarized cross section}

In this subsection, we derive the hard factors $H_{ab\to cd}^{uu}$
for the unpolarized differential cross section in Eq.~(\ref{e2}).
We will first give a simple example of how the hard factors may be
extracted in the context of the quark-diquark model of
\cite{BroHwaSch02}.  We will also give a more general approach
based on power counting techniques to calculate the hard factors
directly from partonic diagrams.

To specify the quark-diquark model of the proton, we give in
Fig.~\ref{f2} graphical rules for the vertex coupling the
quark-diquark to the proton, and the vertex coupling the diquark
to a gauge boson. In order to correctly take into account the
color degree of freedom of the strong interactions, we
included proper color factors for these vertices. Using this
quark-diquark model to extract the hard factors not only verifies
the fact that the hard factors are independent of the dynamics at
the scale of the proton, but also provides a relatively simpler
demonstration of how to factorize the cross sections into the TMD
quark distributions and corresponding hard factors at the leading
order in the $q_\perp/P_\perp$ expansion.

\begin{figure}[t]
\begin{center}
\includegraphics[width=5.5cm]{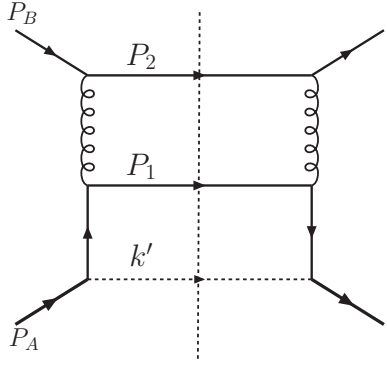}
\end{center}
\vskip -0.4cm \caption{\it The Feynman diagram contributing to the
dijet-correlation in the $qq'\to qq'$ channel, for the unpolarized
cross section. }\label{f3}
\end{figure}

In this particular model, we calculate the cross section for an
energetic parton of momentum $P_B=(0^+,P_B^-,0_\perp)$ scattering
off a parton of momentum $P_A=(P_A^+,0,0_\perp)$ from the (modeled)
nucleon, producing two jets of momenta $P_1$ and $P_2$ plus an unobserved
particle with momentum $k'$ in the final state, as shown in
Fig.~\ref{f3}, where the Feynman diagram represents the
contribution from a quark-quark scattering channel: $qq'\to qq'$
with the quark $q$ from the nucleon. We extract the hard factors
by comparing the calculated cross section with the factorized
formulas in either Eq.~(\ref{e4}) or (\ref{e2}).  The diquark in
the final state has a transverse momentum
$\vec{k}_{\perp}'=-(\vec{P}_{1\perp}+\vec{P}_{2\perp})$ to balance
the small momentum imbalance of the two jets,
$\vec{q}_{\perp}=\vec{P}_{1\perp}+\vec{P}_{2\perp}$.

According to the factorization formulas in
Eqs.~(\ref{e4}),(\ref{e2}), the transverse momentum imbalance $q_\perp$
of the two jets is a consequence of adding the transverse momenta of
the two active partons and from the soft factor. Since the
incoming quark in this model calculation has only a longitudinal
momentum component, its transverse as well as longitudinal
momentum distributions should be given by delta functions,
\begin{equation}
f_b(x_b,P_{B\perp})=\delta(x_b-1) \delta^{(2)}( P_{B\perp})\ .
\end{equation}
Similarly, at this order, the soft factor is also a delta function
of the transverse momentum, i.e.,
$S(\lambda_{\perp})=\delta^{(2)}(\lambda_\perp)$. Therefore, the
only contribution to the transverse momentum imbalance of the two
jets comes from the transverse momentum of the quark emerging from the
nucleon, and the generalized factorization formulas can be reduced
to
\begin{eqnarray}
\frac{d\sigma^{uu}}{dy_1dy_2dP_\perp^2d^2\vec{q}_\perp} &=&
x_af_a(x_a,q_{\perp})H_{ab\to cd}^{uu}(P_\perp^2)\delta(x_b-1)
 \ ,\label{e7}
\end{eqnarray}
for the unpolarized cross section, and
\begin{eqnarray}
\frac{d\Delta\sigma(S_\perp)}{dy_1dy_2dP_\perp^2d^2\vec{q}_\perp}
&=&\frac{\epsilon^{\alpha\beta}S_\perp^\alpha q_\perp^\beta}{M_P}
x_aq_{Ta}^{\rm SIDIS}(x_a,q_{\perp}) H_{ab\to cd}^{\rm
Sivers}(P_\perp^2)\delta(x_b-1)
 \ ,
\label{e8}
\end{eqnarray}
for the single transverse-spin dependent cross section.

\begin{figure}[t]
\begin{center}
\includegraphics[width=9cm]{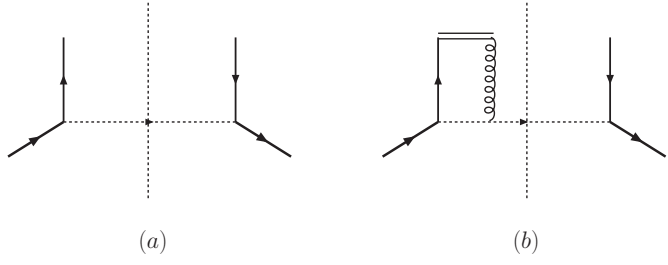}
\end{center}
\vskip -0.4cm \caption{\it  The leading order unpolarized quark
distribution (a) and Sivers function (b), as given in the
quark-diquark model \cite{BroHwaSch02}.  }\label{f4}
\end{figure}

In order to extract the hard factor from the differential cross
sections in Eqs.~(\ref{e7}),(\ref{e8}), we will need to calculate
the TMD quark (unpolarized and Sivers) distributions within the
same model. They are given by the diagrams shown in Fig.~\ref{f4},
and are available \cite{BroHwaSch02,BelJiYua02}. We can summarize
their results as follows:
\begin{eqnarray}
f(x,k_\perp)&=&\frac{g_N^2}{16\pi^3}\frac{1-x}{\Lambda^4(x,k_\perp^2)}
\left[k_\perp^2+(xM_p +m_q)^2\right] \ ,\label{e9}\\
q_T^{\rm
SIDIS}(x,k_\perp)&=&C_F\frac{g_N^2\alpha_s}{16\pi^3}\frac{(1-x)(m_q+xM_p)}
{\Lambda^2(x,k_\perp^2)}\frac{M_p}{k_\perp^2}
\ln\frac{\Lambda^2(x,k_\perp^2)}{\Lambda^2(x,0)} \ , \label{e10}
\end{eqnarray}
where $C_F=(N_c^2-1)/2N_c$ and
$\Lambda^2(x,k_\perp^2)=k_\perp^2+x\lambda_g^2+(1-x)m_q^2-x(1-x)M_p^2$,
and $M_p$, $m_q$ and $\lambda_g$ are masses for the proton, the
quark and the gauge boson, respectively. In the above formulas, we
have included the color-factors arising from the vertices described
in Fig.~\ref{f2}.

We now present details of the calculation of the contribution of
the quark-quark scattering channel $qq'\to q q'$ to the unpolarized
scattering cross section. We also provide full results for all other
channels, which can be derived similarly. The Feynman diagram for
$qq'\to q q'$ has
been shown in Fig.~\ref{f3}. In the limit $q_\perp\ll P_\perp$,
the two final-state jets have approximately the same transverse
momentum, i.e., $P_{1\perp}\approx P_{2\perp}\approx P_\perp$. We
define the following kinematic variables:
\begin{eqnarray}
x_{a1}=\frac{P_\perp}{\sqrt{s}}e^{y_1},~~~
x_{b1}=\frac{P_\perp}{\sqrt{s}}e^{-y_1}\ ,\\
x_{a2}=\frac{P_\perp}{\sqrt{s}}e^{y_2},~~~
x_{b2}=\frac{P_\perp}{\sqrt{s}}e^{-y_2}\ ,
\end{eqnarray}
where $s=(P_A+P_B)^2$. From these, we immediately find the
incident partons' momentum fractions $x_a=x_{a1}+x_{a2}$ and $x_b=
x_{b1}+ x_{b2}$, and the scalar diquark's momentum can be written
as
\begin{equation}
k'=(1-x_a)P_A^++(1-x_b)P_B^-+k_\perp' \ .
\end{equation}
In the evaluation of the cross sections, we use the power counting
analysis \cite{ColSopSte89,sterman-fac}, keep only the leading
power contributions and neglect all higher order corrections in
$\kappa/P_\perp$, where $\kappa$ represents any lower mass scale
like $q_\perp$, $M_p$, $m_q$ and $\lambda_g$. In this limit, the
differential cross section for Fig.~\ref{f3} can be written as
\begin{equation}
\frac{d\sigma^{uu}}{dy_1dy_2dP_\perp^2d^2\vec{q}_\perp}
=\frac{2\pi^2}{2s} \left(\frac{1}{16\pi^3}\right)^2
|\overline{{\cal M}}|^2\delta ((k')^{2}-\lambda_g^2) \ ,
\end{equation}
where ${\cal M}$ is the scattering amplitude for the diagram.
Expanding the above delta function for the phase space integral of
$k'$, we find that in the leading power contributions
\cite{Soperdis},
\begin{equation}
\delta((k')^{2}-\lambda_g^2) =\frac{1}{s}\left\{\frac{
\delta(x_b-1)}{(1-x_a)_+}+\frac{\delta(x_a-1)}{(1-x_b)_+}+\delta
(x_a-1)\delta(x_b-1)\ln\frac{s}{\vec{q}_\perp^2+\lambda_g^2}\right\}
\ , \label{e15}
\end{equation}
where the ``plus'' distribution follows the usual
definition \cite{plus}. This
delta function will help to simplify our calculations for
contributions from different kinematic regions of $k'$. For
example, if $k'$ is parallel to $P_A$, which means $(1-x_a)\neq
0$, we only have a contribution from the first term of the above
expansion, which also implies $x_b=1$ in this limit. Furthermore,
the quark propagator in Fig.~\ref{f3} reads
\begin{equation}
\frac{1}{(P_A-k')^2-m_q^2}=-\frac{1-x_a}{\Lambda^2(x_a,q_\perp^2)}
\ ,
\end{equation}
where $\Lambda^2(x_a,q_\perp^2)$ is defined above after
Eq.~(\ref{e10}), and is order of $\vec{q}_\perp^2$. Combining this
propagator with the delta function expansion, we find that indeed
only the first term in the delta function contributes.
Neglecting all higher order terms in $\kappa/P_\perp$, we obtain
the amplitude squared for Fig.~\ref{f3} as
\begin{equation}
|\overline{\cal M}|^2 =
\frac{N_c^2-1}{4N_c^2}\left(\frac{-2}{t^2}\right)
\frac{g_N^2\alpha_s^2(4\pi)^2}{((P_A-k')^2-m_q^2)^2} \left(x_a s t
+2 s u + tu \right) \left[(x_aM_p+m_q)^2+\vec{q}_\perp^2\right] \
,
\end{equation}
where $({N_c^2-1})/{4N_c^2}$ is the color-factor, and the hadronic
Mandelstam variables $s$, $t$, and $u$ are defined as
$s=(P_A+P_B)^2$, $t=(P_A-P_1)^2$, $u=(P_B-P_1)^2$. In the limit
$x_b=1$, these variables can be related to the partonic Mandelstam
variables as: $s\approx \hat s/x_a$, $t\approx \hat t/x_a$,
$u\approx \hat u$ at ${\cal O}(P_\perp^2)$, and we will have $\hat
s+\hat t+\hat u=0$. Substituting the above results into the
differential cross section, we obtain,
\begin{equation}
\frac{d\sigma^{uu}}{dy_1dy_2dP_\perp^2d^2\vec{q}_\perp} =
\frac{N_c^2-1}{4N_c^2} \left[ \frac{g_N^2}{16\pi^3}\, x_a(1-x_a)
\frac{(x_aM_p+m_q)^2+\vec{q}_\perp^2}{\Lambda^4(x_a,q_\perp^2)}
\right] \frac{2\pi\alpha_s^2}{\hat s^2}\, \frac{\hat s^2+\hat
u^2}{\hat t^2}\, \delta(x_b-1) \ .
\end{equation}
Comparing this result with Eq.~(\ref{e7}), and identifying the
TMD parton distribution $x_a f(x_a,q_\perp)$ defined in
Eq.~(\ref{e9}), we extract the partonic hard factor for the
differential cross section from the diagram in Fig.~3
as\begin{equation} H_{qq'\to qq'}^{uu}=\frac{\alpha_s^2\pi}{\hat
s^2}\frac{N_c^2-1}{4N_c^2}\frac{2(\hat s^2+\hat u^2)}{\hat t^2} \
, \label{hqqp}
\end{equation}
where $\hat s,~\hat t,~\hat u$ are the partonic Mandelstam
variables defined before. This hard factor is just the well-known
differential cross section $d\hat \sigma/d\hat t$ for the partonic
process $qq'\to qq'$. The hard factors for all other partonic
channels can be calculated similarly, and they are all identical
to the partonic differential cross section $d\hat\sigma/d\hat t$,
and collected in Appendix A.

Instead of calculating the differential cross section directly, we
now show how to use the power counting techniques
\cite{ColSopSte89} to analyze the contribution from all partonic
channels and to demonstrate the factorization of the hard factors.
We then apply this approach to calculate the contributions to the
spin dependent differential cross section.

We again first consider the simple example of the $qq'\to qq'$ channel.
The jet transverse momentum $P_\perp$ is the only hard
scale, while all other momentum scales, such as the imbalance $q_\perp$
of the two jets and all other mass scales $M_p$, $m_q$ and
$\lambda_g$, are relatively soft and much smaller than the hard
scale. We expand the scattering amplitude squared in the small
parameter $q_\perp/P_\perp$. Using power counting techniques,
we identify the leading power contributions and match them to the
momentum regions of the unobserved (or integrated) parton of
momentum $k'$. In our current model, where a parton of momentum $P_B$
scatters off a nucleon of momentum $P_A$, the leading power
contribution is from the region when $k'$ is parallel to $P_A$. If
$k'$ is parallel to $P_B$, the contribution is power suppressed,
because the denominator of the quark propagator $(P_A-k')^2-m_q^2$
will be of order $s=(P_A+P_B)^2\sim P_\perp^2$, instead of
${\cal O}(\vec{q}_\perp^2)$ as is the case
when $k'$ is parallel to $P_A$.

\begin{figure}[t]
\begin{center}
\includegraphics[width=13cm]{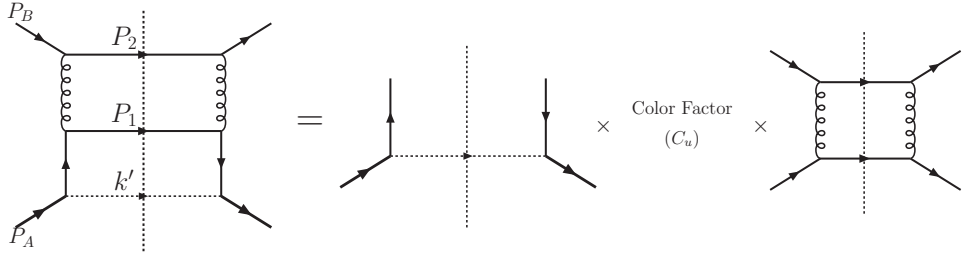}
\end{center}
\vskip -0.4cm \caption{\it Factorization of the unpolarized cross
section of Fig.~\ref{f3} into the TMD quark distribution of the proton
and the hard factor represented by the partonic
scattering Feynman diagram.}\label{f5}
\end{figure}

When the momentum $k'$ is parallel to $P_A$, the quark propagator
in Fig.~\ref{f3} has a virtuality $\Lambda^2(x_a,q_\perp^2)\sim
q_\perp^2$, which is much smaller than the hard scale $P_\perp^2$.
Compared to the short distance physics in the partonic scattering
taking place at the time scale $1/P_\perp$, the incident quark of
momentum $P_A-k'$ is long-lived, and the contribution from this
subprocess can be separated into two parts: the parton
distribution part relevant at the soft scale $\sim {\cal O}(q_\perp)$ and
the hard partonic part at the scale $\sim {\cal O}(P_\perp)$,
representing the long distance and short distance physics,
respectively.  We demonstrate this factorization in Fig.~\ref{f5},
where the contribution to the differential cross section from the
process in Fig.~\ref{f3} is factorized into a parton distribution
multiplied by a hard factor. We can further separate the hard factor
into a partonic scattering amplitude squared stripped of any color,
and the color-factor $C_u={\rm Tr}(T^aT^b){\rm Tr}(T^aT^b)/N_c^2=
(N_c^2-1)/4N_c^2$. In this way we can take into account the color
decomposition of the subprocess in the case of unpolarized scattering.
This is trivial for this particularly simple partonic channel, but
becomes a convenience for the more complicated ones.
After subtraction of the parton distribution from the differential
cross section, the hard factor in Fig.~\ref{f5} can then be written as
\begin{equation}
H_{qq'\to qq'}^{uu}=\frac{\alpha_s^2\pi}{\hat s^2}C_u\times
h_{qq'\to qq'} \ ,\label{hqu}
\end{equation}
where $\alpha_s^2\pi/\hat s^2$ represents a common factor and is
introduced to simplify the notation, and $h_{qq'\to qq'}=2(\hat s^2+\hat
u^2)/\hat t^2$. This of course agrees with the above result in
Eq.~(\ref{hqqp}). Most importantly, it demonstrates that the
hard factor is model independent. It only depends on the partonic
scattering cross section and the color-factor
associated with the partonic diagram. It does not depend on the
model we used to describe how the nucleon couples to the quark and
diquark.

\subsection{Hard factors for the spin-dependent cross section}

\begin{figure}[t]
\begin{center}
\includegraphics[width=10cm]{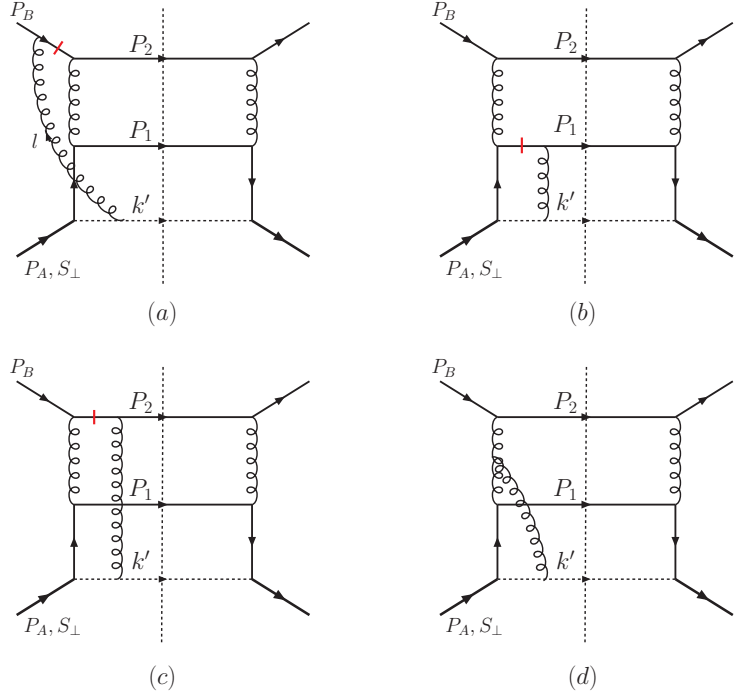}
\end{center}
\vskip -0.4cm \caption{\it Initial and final state interactions
between the scattering partons and the diquark of the polarized
proton: (a) for initial state interaction, and (b,c) for the final
state interactions diagrams, and (d) for interaction with the
internal gluon propagator which does not contribute to the soft
gluon pole corresponding to the quark Sivers function contribution
to the dijet-correlation SSA.}\label{f6}
\end{figure}

A non-vanishing single transverse-spin asymmetry requires
initial/final state interactions generating a phase. Because all
initial and final partonic states relevant
for dijet production are colored, both
initial and final state interactions have to be taken into account
for a complete result. In Fig.~\ref{f6}, we show all possible
diagrams with gluon exchange between the diquark and the hard
scattering part which may contribute to the SSA for the
dijet-correlation, again for the case of an underlying $qq'\to qq'$
process. The initial and final state interaction
diagrams in Fig.~\ref{f6}(a-c) contribute to a soft gluon pole,
which corresponds to the quark Sivers function contribution we are
considering in this paper. However, the gluon interaction with the
internal gluon propagator shown in Fig.~\ref{f6}(d) does not
contribute to a soft gluon pole, but to a soft fermion pole
\cite{qiusterman}. This soft fermion pole contribution is not
related to the quark Sivers function and will not be discussed in
this paper. Therefore, in the following calculations, we will only
consider the contributions from the diagrams (a)-(c).

We will follow the same factorization approach that was used in
last subsection for calculating the hard factors for the
unpolarized cross section. We use the power counting method to
factorize the additional gluon interaction between the hard
scattering and the diquark
into the Sivers parton distribution for every partonic channel. As
discussed above, the leading power contribution is from the region
of phase space where the outgoing diquark momentum $k'$ is
parallel to the polarized proton $P_A$: $k^{\prime +}\sim P_A^+$,
$k^{\prime -}\sim q_\perp^2/P_A^+$ and $k'_\perp\sim q_\perp$. To
obtain a leading power contribution, the gluon momentum $l$ also needs
to be parallel to $P_A$. This is so because the
diquark propagator in Fig.~\ref{f6} reads
\begin{equation}
\frac{1}{(k'+l)^2+i\epsilon}\approx \frac{1}{2k^{\prime
+}l^-+{\cal O}(q_\perp^2)} \ ,
\end{equation}
which will be power suppressed unless $l^-\sim q_\perp^2/P_A^+$.
This result can also be derived from the existence of a
``pinch'' singularity in the $l^-$ integral over the quark propagator
of momentum $P_A-k'-l$ and the diquark propagator of momentum
$l+k'$ \cite{bodwin}. Since the momentum $l$ of the gluon is parallel to
$P_A$, its polarization will also be proportional to $P_A$, because
the gauge-boson-scalar vertex is proportional to $(2k'-l)^\mu$, as
shown in Fig.~\ref{f2}. Therefore, the gluon
attaching to the initial- or final-state quark is
longitudinally polarized, i.e., its polarization is along its
momentum. Because of this, we can further decouple the gluon
interaction with the external particles by using the eikonal
approximation. For example, for the diagram with final-state
interaction in Fig.~\ref{f6}(b), the gluon interaction part can be
reduced to
\begin{eqnarray}
\bar u(P_1)T^a(-ig)\gamma^- \frac{i(\not\!P_1-\not\!
l+m_q)}{(P_1-l)^2-m_q^2+i\epsilon} \dots &\approx &
\frac{2P_1^-g}{-2P_1^-\cdot l^++i\epsilon}\bar u(P_1)T^a\dots\nonumber\\
&=&\frac{g}{-l^++i\epsilon}\bar u(P_1)T^a\dots \ ,
\end{eqnarray}
where the Gamma matrix $\gamma^-$ appears because of the interaction
with a longitudinally polarized gluon, and where
$a$ is the color-index for this gluon. In the above derivations,
we have only kept the leading power contributions.

\begin{figure}[t]
\begin{center}
\includegraphics[width=14cm]{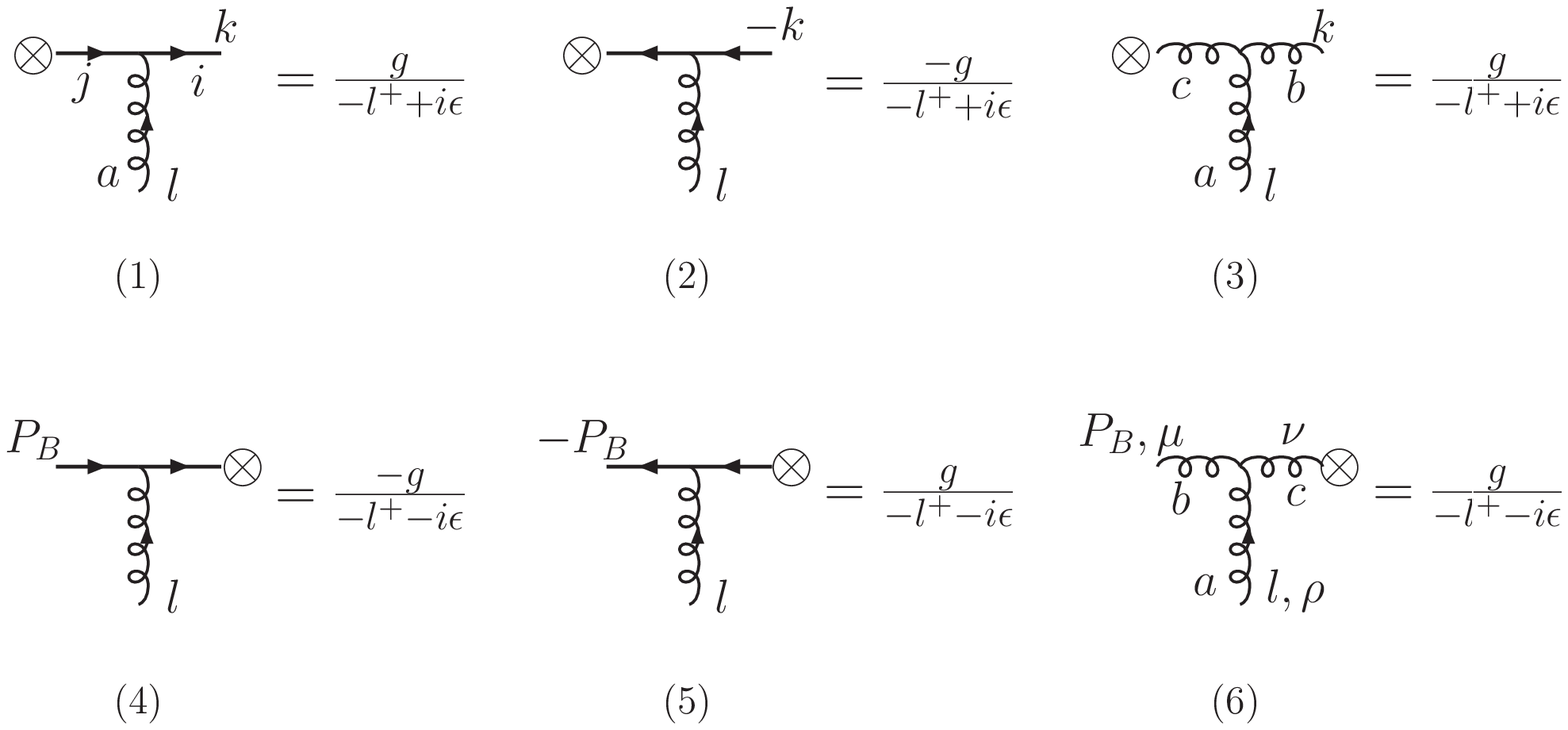}
\end{center}
\vskip -0.4cm \caption{\it Eikonal approximation for the attachment of
a longitudinally polarized gluon to external lines: (1-3) for the
final state interactions when the gluon attaches to outgoing
particles; (4-6) for the initial state interactions for incoming
particles. The color-factor is not shown in the above and should
read: $T^a_{ij}$ for (1,2,4,5) where $a$ is the color index for
the gluon with momentum $l$ and $i,j$ are indices for the quark
lines; $(-if_{abc})$ for (3,6) where $b,c$ are indices for
incoming and outgoing gluons.}\label{f7}
\end{figure}

Similarly, the initial state interaction in Fig.~\ref{f6}(a) and
the other final state interaction in Fig.~\ref{f6}(c) can be
simplified following the same eikonal approximation. The eikonal
approximation will give rise to eikonal propagators and vertices.
In Fig.~\ref{f7}, we summarize these eikonalized expressions for
all the initial and final state interactions, including the
attachments of the longitudinally polarized gluon to a quark or
anti-quark line, or to a gluon line as well. The eikonal
approximation for quark- and antiquark-gluon interactions with
incoming or outgoing momentum can be derived following the above
example. For the three-gluon interactions, we can derive the
eikonal vertex and propagator by choosing physical
polarizations for the external gluons. With this choice, we find
the vertex and gluon propagator can be reduced to, for example for
Fig.~\ref{f7}(6):
\begin{eqnarray}
(-ig)(-if_{abc})\!\!&\!\!&\left[(l-P_B)_{\nu'}
g_{\rho\mu}+(2P_B+l)_\rho g_{\mu\nu'}+(-2l-P_B)_\mu
g_{\nu'\rho}\right]\frac{i(-g^{\nu\nu'})}{(P_B+l)^2+i\epsilon}g^{\rho
-}\nonumber\\&&\approx  \frac{(2P_B+l)^-
g_{\mu\nu}}{(P_B+l)^2+i\epsilon}(-if_{abc})=\frac{g}{-l^+-i\epsilon}
g_{\mu\nu}(-if_{abc}) \ ,
\end{eqnarray}
where we have used the fact that the incoming gluon is
transversely polarized, so that the polarization tensor can
be chosen as
\begin{equation} \label{glupol}
\sum_\lambda
\epsilon^\mu_\lambda(P_B)\epsilon^{\mu'*}_\lambda(P_B)
=-g^{\mu\mu'}+\frac{P_B^\mu P_A^{\mu'}
                   +P_B^{\mu'} P_A^{\mu}}{P_A\cdot P_B} \ .
\end{equation}
Similarly, we can simplify the longitudinally polarized gluon
attachment to the outgoing gluon line as shown in
Fig.~\ref{f7}(3).

\begin{figure}[t]
\begin{center}
\includegraphics[width=14cm]{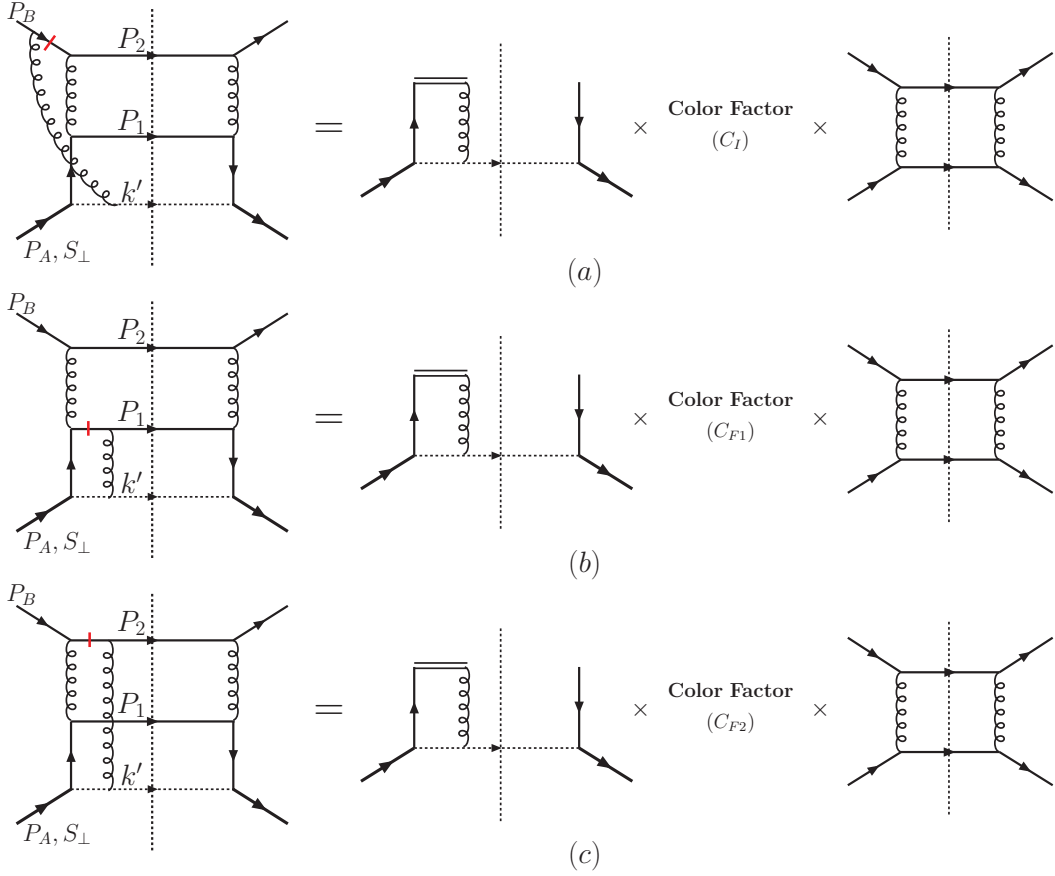}
\end{center}
\vskip -0.4cm \caption{\it Factorization of the initial/final state
interaction contributions to the SSA in dijet correlations into the
quark Sivers function (defined as in SIDIS), multiplied in each case
by an adjusted color-factor and a squared scattering amplitude for
the partonic process without color factor, similar to
Fig.~\ref{f5}. } \label{f8}
\end{figure}

These eikonal propagators and the associated vertices can be
absorbed into the parton distribution for the polarized proton,
after factorizing the diagram. The only difference between this
factorization and the one for the unpolarized scattering is the
color-flow, because of the additional gluon exchange between the
diquark and the hard part. The color-flow is sensitive to the
difference between initial- and final-state interactions and
depends on the specific partonic subprocess.  Therefore, the
generalized factorization would immediately fail if the eikonal
propagators and associated vertices, which are absorbed into the TMD
distributions, were sensitive to the process dependent color flows.
However, at the leading order, one can always attribute the
color-flow to the partonic hard part and keep the corresponding
TMD distribution insensitive to it. Therefore, as compared to
the unpolarized case (see Fig.~\ref{f5}), for the single-spin-dependent
cross section we shall obtain a more complicated factorization.
For the case of the three diagrams of Fig.~\ref{f6} the
factorization is shown in Fig.~\ref{f8}.
The contribution from each diagram can be factorized
into the Sivers function and the hard partonic scattering
amplitude squared, multiplied by a color-factor that takes
into account the color-flow of the initial/final state interaction.
However, in order to verify the generalized factorization beyond the
leading order, one would have to show that additional initial-state
as well as final-state gluon interactions could also be factorized
into a sum of terms, each of which is factorized into a product of
higher order TMD distributions and corresponding color factors and
partonic hard parts. For example, to verify the factorization at
the next-to-leading order with two eikonal gluon interactions, one
would have to show that the initial- and final-state interactions
can be factorized into the same color factors and the lowest order
hard parts as in Fig.~8,
multiplied by the corresponding second order TMD distribution. We
will leave the detailed study of the generalized factorization at
higher orders to a future publication.

At least to the leading order, the extracted color factors
$C_{i}$ with $i=I,F_1,F_2$ in Fig.~\ref{f8} are insensitive to the
long-distance physics. That is, they do not depend on how the
quarks and the gluon couple to the nucleon. This is because the
factorization of the color-flow between the hard partonic part and
the TMD part results in a color configuration connecting
an incoming and an outgoing quark (with color indices $ij$)
and a gluon (with color $a$). The only possible color structure
to describe this
connection is $T^a_{ij}$, due to $SU(3)$ invariance and the
Wigner-Eckart theorem. Therefore, one can reduce all the color matrices
from the hard part to a simple color matrix $T^a_{ij}$, and the
coefficient will be exactly the color-factor in the factorization
formula. In this way, the extracted color-factor
cannot depend on the bottom part of the diagram describing the
coupling of the quarks and the gluon to the nucleon, but will only
depend on the color-flow in the
hard partonic diagrams with an additional gluon insertion. For
example, the color factors appearing in Fig.~\ref{f8} can be
derived in the following way,
\begin{eqnarray}
{\rm \ref{f8}(a)}&:& \frac{1}{N_c}{\rm
Tr}(T^bT^cT^a)(T^bT^c)_{ij}=\frac{-1}{2N_c^2}T^a_{ij}
\Longrightarrow C_I=\frac{-1}{2N_c^2}\ , \label{e25}\\
{\rm \ref{f8}(b)}&:& \frac{1}{N_c}{\rm
Tr}(T^bT^c)(T^bT^aT^c)_{ij}=\frac{-1}{4N_c^2}T^a_{ij}
\Longrightarrow C_{F1}=\frac{-1}{4N_c^2}\ , \label{e26} \\
{\rm \ref{f8}(c)}&:& \frac{1}{N_c}{\rm
Tr}(T^bT^aT^c)(T^bT^c)_{ij}=\frac{N_c^2-2}{4N_c^2}T^a_{ij}
\Longrightarrow C_{F2}=\frac{N_c^2-2}{4N_c^2}\ ,  \label{e27}
\end{eqnarray}
where the color-matrices on the left side of the equations come
from the partonic diagrams with one gluon attachment to the
initial- or final-state quark line. The extracted color-factor
$C_I$ is for the initial-state interaction, and $C_{F1}$ and
$C_{F2}$ are for the final-state interactions on the lines
with the momentum $P_1$ and $P_2$, respectively.

In the factorization shown in Fig.~\ref{f8}, the leading order
Sivers functions for all three diagrams are the same. This is
because we normalize all eikonal vertices to the one used in the
calculations of the Sivers function in SIDIS (shown in
Fig.~\ref{f4}). The difference from the initial or final state
interaction effects is summarized into the relevant
color-factors. For other partonic channels, it is in some cases
necessary to introduce an extra sign when the eikonal propagator
contributes with an opposite phase compared to that in the Sivers
function in SIDIS. For example, the eikonal
propagators in the diagrams (2,5,6) of Fig.~\ref{f7} contribute an
opposite sign, whereas those of (1,3,4) contribute the same sign as
in SIDIS. Because all the initial and final state interactions
in Fig.~\ref{f8} contribute the same sign, there is no sign change
for their color-factors.

Another important point is that the partonic scattering amplitude
squared is the same for all the three diagrams in Fig.~\ref{f8}.
It can be calculated directly from the relevant Feynman rules for
the partonic diagrams and is also the same as the one for
unpolarized scattering given above. After factorizing out the
Sivers function from the cross section contributions for the
diagrams in Fig.~\ref{f8}, the color factors for the three diagrams
can then be summed to give the final result for the hard factor
for the $qq'\to qq'$ subprocess:
\begin{equation}
H_{qq'\to qq'}^{\rm Sivers}=\frac{\alpha_s^2\pi}{\hat
s^2}\left[C_I+C_{F1}+C_{F2}\right]h_{qq'\to qq'}=
\frac{\alpha_s^2\pi}{\hat s^2}\frac{N_c^2-5}{4N_c^2}\frac{2(\hat
s^2+\hat u^2)}{\hat t^2} \ , \label{hqq}
\end{equation}
where $h_{qq'\to qq'}$ has been defined above after Eq.~(\ref{hqu}).

The above derivations can be extended to all other partonic
channels involving the quark Sivers function. They can
also be extended to the case of sub-processes initiated by the
gluon Sivers function for which, however, more color-structures (like
$f_{abc}$ and $d_{abc}$ for the three gluon coupling to the
nucleon state) will emerge \cite{ji-gluon}. In this paper, we will
focus on the processes with a quark Sivers function, which include
all quark-quark and quark-antiquark scattering processes, such as
$qq'\to qq'$ calculated above,
$q\bar q'\to q\bar q'$, $q \bar q\to q'\bar q'$, $qq\to qq$ and
$q\bar q\to q\bar q$, as well as the quark-gluon process $qg\to qg$ and
quark-antiquark annihilation into a gluon pair, $q\bar q\to gg$.
Following the above derivations for the $qq'\to qq'$ process, the
hard factors for the unpolarized and single-spin-dependent cross
sections for these channels can be summarized by the following
master formulas:
\begin{eqnarray}
H_{ab\to cd}^{uu} &=&\frac{\alpha_s^2\pi}{\hat s^2}\sum_i C_u^i
h^i_{ab\to cd}
\label{e29}\\
H_{ab\to cd}^{\rm Sivers}&=&\frac{\alpha_s^2\pi}{\hat s^2}\sum_i
(C_I^i+C_{F1}^i+C_{F2}^i) h^i_{ab\to cd} \ , \label{e30}
\end{eqnarray}
where $i$ labels an individual Feynman diagram (meant as the square of an
amplitude or an interference between two amplitudes, see below)
with $h^i$ the associated expression for the hard factor. $C_u^i$ is the
color-factor for the unpolarized cross section, $C_I^i$ for the initial state
interaction for the single-spin dependent cross section, and
$C_{F1}^i$ and $C_{F2}^i$ for the final state interactions when the
gluon attachment is to the lines of momentum $P_1$ and $P_2$ respectively.

From the above analysis, we can easily see that the hard factors in
Eqs.~(\ref{e29}),(\ref{e30}) are model-independent. The
model-dependence is removed once we factorize out the unpolarized
quark distribution or the Sivers function from the differential
cross sections. From the above master formulas, the hard factors
depend on the partonic scattering amplitude squared, and the
color-factors, derived from partonic diagrams with or without
the additional gluon attachment. Both the color and partonic
factors, and therefore the hard factors, are independent of the model
for the proton used in our calculation.

\begin{figure}[t]
\begin{center}
\includegraphics[width=13cm]{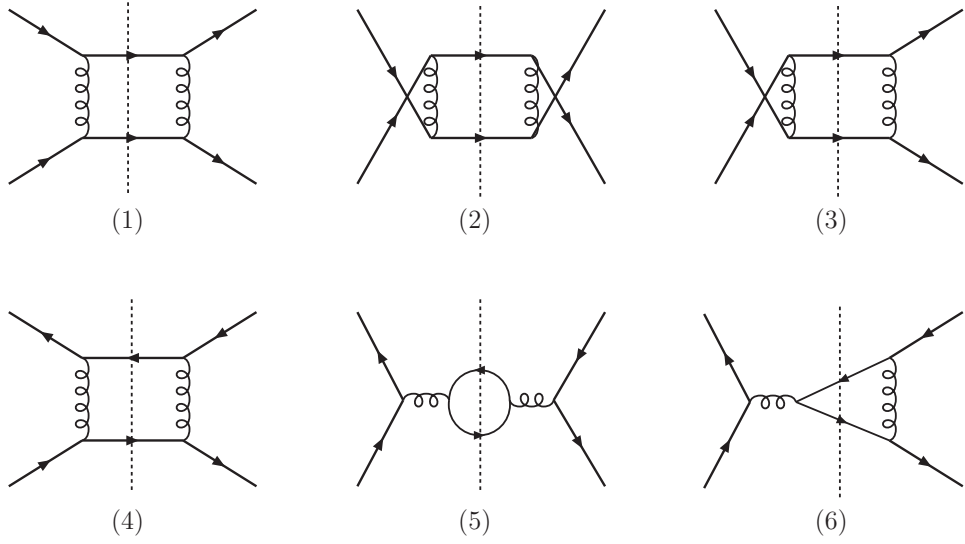}
\end{center}
\vskip -0.4cm \caption{\it Quark-(anti)quark scattering diagrams.
The mirror diagrams of (3) and (6) are not shown; their
contributions in Table I are identical.} \label{f9}
\end{figure}
\begin{table}[t]
\caption{The color-factors and hard cross sections of individual
diagrams of Fig.~\ref{f9} for the quark-quark scattering channels.
}
\begin{ruledtabular}
\begin{tabular}{|l|c|c|c|c|c|c|}
& (1) & (2) & (3) & (4) & (5) & (6)  \\
\hline $h$ ~~~& $~~\frac{2(\hat s^2+\hat u^2)}{\hat t^2}~~$
&$~~\frac{2(\hat s^2+\hat t^2)}{\hat u^2}~~$ &$~~\frac{2\hat
s^2}{\hat t\hat u}~~$ & $~~\frac{2(\hat s^2+\hat u^2)}{\hat
t^2}~~$ & $~~\frac{2(\hat t^2+\hat u^2)}{\hat s^2}~~$
&$~~\frac{2\hat u^2}{\hat s\hat t}~~$\\
\hline $C_u$  & $\frac{N_c^2-1}{4N_c^2}$ &$\frac{N_c^2-1}{4N_c^2}$
       &$-\frac{N_c^2-1}{4N_c^3}$ & $\frac{N_c^2-1}{4N_c^2}$&
$\frac{N_c^2-1}{4N_c^2}$& $-\frac{N_c^2-1}{4N_c^3}$\\
\hline $C_{I}$ & $-\frac{1}{2N_c^2}$ & $-\frac{1}{2N_c^2}$
       & $\frac{N_c^2+1}{4N_c^3}$ &$-\frac{N_c^2-2}{4N_c^2}$ &
$\frac{1}{4N_c^2}$&$-\frac{1}{4N_c^3}$\\
\hline $C_{F1}$ & $-\frac{1}{4N_c^2}$ & $\frac{N_c^2-2}{4N_c^2}$
& $\frac{1}{4N_c^3}$& $-\frac{1}{4N_c^2}$& $\frac{N_c^2-2}{4N_c^2}$&
$\frac{1}{4N_c^3}$\\
\hline $C_{F2}$ & $\frac{N_c^2-2}{4N_c^2}$ &$-\frac{1}{4N_c^2}$ &
$\frac{1}{4N_c^3}$ & $\frac{1}{2N_c^2}$& $\frac{1}{2N_c^2}$&
$-\frac{N_c^2+1}{4N_c^3}$
 \end{tabular}
\end{ruledtabular}
\end{table}

For the quark-(anti)quark scattering channels, we plot all the
relevant diagrams in Fig.~\ref{f9}. The color-factors for
unpolarized scattering can be straightforwardly evaluated. In
the spin-depedent case, the color-factors for each partonic channel
can be calculated following the above example for $qq'\to qq'$,
by reducing the strings of color-matrices for the partonic diagrams with
an additional initial- or final-state gluon attachment to the
simple $T^a_{ij}$, and extracting the coefficients as the
relevant color-factors. We list all these color-factors in Table I,
including those for the unpolarized scattering. The color-factors in
the spin-dependent case can also be calculated by contracting
the string of color-matrices for a given partonic diagram with
a $T^a$, analogous to what was done in \cite{qiusterman}. These two
methods yield the same answer. The partonic amplitude squared for
each diagram in Fig.~\ref{f9} can be easily calculated using the
Feynman rules in Feynman gauge. We also list the results for the
$h^i$ in Table I. Combining the results given in Table I,
we can obtain the hard factor for any particular quark-(anti)quark
scattering channel. For example, for the $qq\to qq$ subprocess, we
have contributions from diagrams (1-3) of Fig.~\ref{f9}. So, the
hard factor for this process will be
\begin{eqnarray}
H_{qq\to qq}^{\rm Sivers}&=&\frac{\alpha_s^2\pi}{\hat
s^2}\left\{h^{(1)}\left[C_I^{(1)}+C_{F1}^{(1)}+C_{F2}^{(1)}\right]
+h^{(2)}\left[C_I^{(2)}+C_{F1}^{(2)}+C_{F2}^{(2)}\right]
\right.\nonumber\\
&&\left.+2
h^{(3)}\left[C_I^{(3)}+C_{F1}^{(3)}+C_{F2}^{(3)}\right]\right\}\ ,
\end{eqnarray}
where a factor $2$ in the third term comes from the mirror diagram
of $(3)$. The full hard factors for all processes are listed in
the Appendix.

\begin{figure}[t]
\begin{center}
\includegraphics[width=13cm]{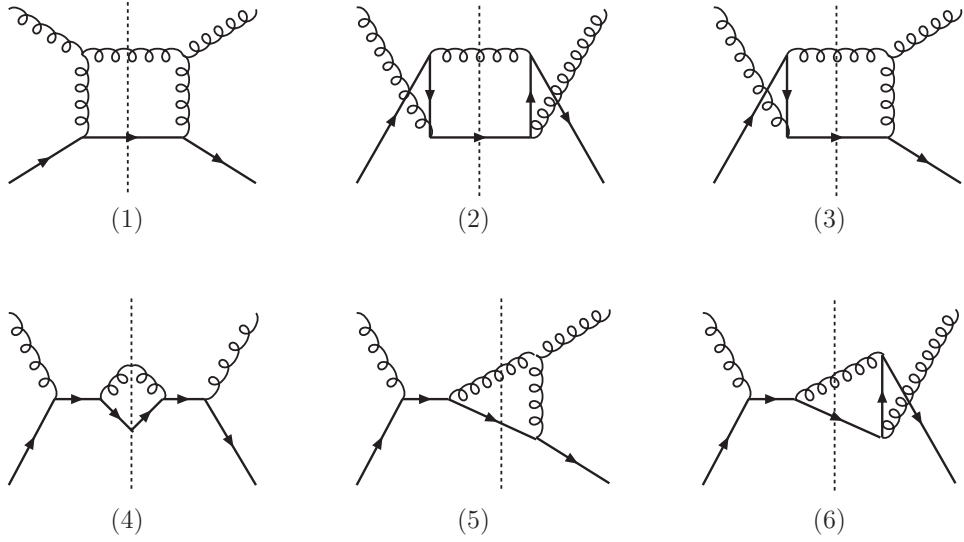}
\end{center}
\vskip -0.4cm \caption{\it Quark-gluon scattering diagrams. The
mirror diagrams of (3), (5) and (6) are not shown; their contributions
in Table II are identical.}\label{f10}
\end{figure}
\begin{table}[t]
\caption{The color- and hard factors for the $qg\rightarrow qg$
scattering channels in Fig.~\ref{f10}, where $C_F=(N_c^2-1)/2N_c$.
}
\begin{ruledtabular}
\begin{tabular}{|l|c|c|c|c|c|c|}
& (1) & (2) & (3) & (4) & (5) & (6)  \\
\hline $h$ ~~~& $~~-\frac{4(\hat t^2-\hat s\hat u)^2}{\hat t^2\hat
s\hat u}~~$ &$~~-\frac{2(\hat u^2+\hat t^2)}{\hat s\hat u}~~$
&$~~\frac{2(\hat t^2-\hat s\hat u)(\hat u-\hat t)}{\hat s\hat
t\hat u}~~$ & $~~-\frac{2(\hat s^2+\hat t^2)}{\hat s\hat u}~~$ &
$~~-\frac{2(\hat t^2-\hat s\hat u)(\hat s-\hat t)}{\hat s\hat
t\hat u}~~$
&$~~\frac{2\hat t^2}{\hat s\hat u}~~$\\
\hline $C_u$  & $\frac{1}{2}$ &$\frac{C_F}{2N_c}$
       &$-\frac{1}{4}$ & $\frac{C_F}{2N_c}$& $\frac{1}{4}$ &
$-\frac{1}{4N_c^2}$\\
\hline $C_{I}$ & $-\frac{N_c^2}{4(N_c^2-1)}$ &
     $\frac{1}{4(N_c^2-1)}$
       & $0$ &$-\frac{1}{4}$ & $-\frac{N_c^2}{4(N_c^2-1)}$&$
\frac{1}{4(N_c^2-1)}$\\
\hline $C_{F1}$ & $-\frac{1}{2(N_c^2-1)}$ &
$\frac{1}{4(N_c^2-1)N_c^2}$
& $\frac{1}{4(N_c^2-1)}$& $\frac{1}{4(N_c^2-1)N_c^2}$&
$-\frac{1}{4(N_c^2-1)}$&$\frac{N_c^2+1}{4(N_c^2-1)N_c^2}$\\
\hline $C_{F2}$ & $\frac{N_c^2}{4(N_c^2-1)}$ &$\frac{1}{4}$ &
$-\frac{N_c^2}{4(N_c^2-1)}$ & $-\frac{1}{4(N_c^2-1)}$& 0&
$-\frac{1}{4(N_c^2-1)}$
 \end{tabular}
\end{ruledtabular}
\end{table}
\begin{figure}[t]
\begin{center}
\includegraphics[width=13cm]{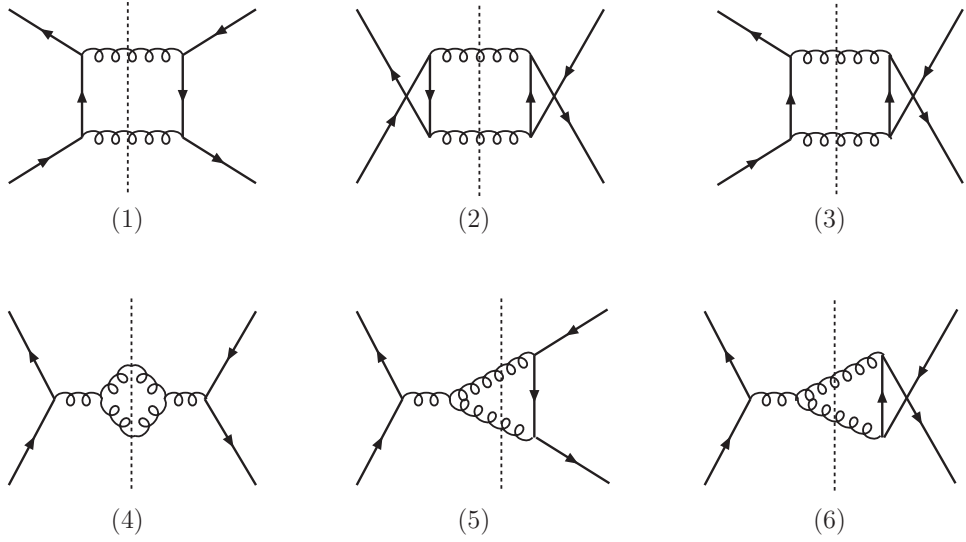}
\end{center}
\vskip -0.4cm \caption{\it The $q \bar q\to gg$ scattering
diagrams. The mirrors of (3), (5) and (6) are not shown; their
contributions in Table III are identical.}\label{f11}
\end{figure}
\begin{table}[t]
\caption{The color- and hard factors for the $q\bar q\rightarrow
gg$ scattering channels in Fig.~\ref{f11}. }
\begin{ruledtabular}
\begin{tabular}{|l|c|c|c|c|c|c|}
& (1) & (2) & (3) & (4) & (5) & (6)  \\
\hline $h$ ~~~& $~~\frac{2(\hat s^2+\hat t^2)}{\hat t\hat u}~~$
&$~~\frac{2(\hat s^2+\hat u^2)}{\hat t\hat u}~~$ & $~~-\frac{2\hat
s^2}{\hat t\hat u}~~$ &$~~\frac{4(\hat s^2-\hat t\hat u)^2}{\hat
s^2\hat t\hat u}~~$  & $~~\frac{2(\hat s^2-\hat t\hat u)(\hat
t-\hat s)}{\hat s\hat t\hat u}~~$ & $~~\frac{2(\hat s^2-\hat t\hat
u)(\hat u-\hat s)}{\hat s\hat
t\hat u}~~$  \\
\hline $C_u$  & $\frac{C_F^2}{N_c}$ &$\frac{C_F^2}{N_c}$
       &$-\frac{C_F}{2N_c^2}$ & $C_F$& $\frac{1}{2}C_F$ & $\frac{1}{2}C_F$\\
\hline $C_{I}$ & $-\frac{1}{4N_c^3}$ &
     $-\frac{1}{4N_c^3}$
       & $-\frac{N_c^2+1}{4N_c^3}$ &$\frac{1}{2N_c}$ & $\frac{1}{4N_c}$&
$\frac{1}{4N_c}$\\
\hline $C_{F1}$ & $\frac{1}{2}C_F$ & $-\frac{1}{4N_c}$
& $-\frac{1}{4N_c}$ & $\frac{N_c}{4}$& $\frac{N_c}{4}$& 0\\
\hline $C_{F2}$ & $-\frac{1}{4N_c}$ &$\frac{1}{2}C_F$ &
$-\frac{1}{4N_c}$ & $\frac{N_c}{4}$& 0& $\frac{N_c}{4}$
 \end{tabular}
\end{ruledtabular}
\end{table}

In Fig.~\ref{f10}, we show the diagrams for the quark-gluon
scattering channel $qg\to qg$.  Our results for this channel are
listed in Table II. The color-factors are calculated following
the method described above. In order to factorize the cross section
and to make the eikonal approximation, as mentioned above, we choose
a physical polarization for the external gluons, using the
polarization projection in Eq.~(\ref{glupol}) for the incoming gluon with
momentum $P_B$, and similarly for an outgoing gluon,
\begin{equation}
\sum_\lambda
\epsilon^\mu_\lambda(k)\epsilon^{\mu'*}_\lambda(k)=-g^{\mu\mu'}+\frac{k^\mu
P_A^{\mu'}+k^{\mu'} P_A^{\mu}}{P_A\cdot k} \ ,
\end{equation}
where $k=P_1$ or $k=P_2$.
With these choices, we calculate the partonic amplitude squared,
$h_{qg\to qg}^i$, for each diagram in Fig.~\ref{f10} and list all results
in Table II.  Again, the full hard factors for this channel can be
derived from our results in Table II using the master formulas in
Eqs.~(\ref{e29}) and (\ref{e30}) for the unpolarized and polarized
scatterings, respectively.

In Fig.~\ref{f11}, we show the relevant
diagrams for the $q\bar q\to gg$ channel.  Our results for the
color-factors and the squared partonic amplitudes are listed in
Table III. The hard factors can be calculated accordingly, and
their results are listed in the Appendix.

We note that our approach for deriving the hard factors for the
dijet-correlation can be straightforwardly extended to the case of
jet-photon correlations in hadronic reactions. This process
was also considered in Ref.~\cite{Bacchetta:2007sz}.
Here we have contributions
from the partonic channels $qg\to q\gamma$ and $q\bar q\to g\gamma$.
The hard factors for both channels can be derived from the above
results by replacing one final-state gluon by a photon. We list the
full corresponding hard factors for both the unpolarized and the
single-spin dependent cross sections, which agree with the ones
of~\cite{Bacchetta:2007sz}, in the Appendix.  It is interesting
to note that the initial state interaction dominates for the $qg\to
q\gamma$ channel, whereas for $q\bar q\to g\gamma$ the final state
interaction dominates.

As the above calculations have demonstrated, at the order we
have considered the TMD parton distributions extracted from the SIDIS
(or Drell-Yan)
process can be used to predict the cross sections or the azimuthal
asymmetries for the dijet-correlation at hadron colliders.
Of course, the initial and final state interactions contributing
to the SSA in dijet-correlations will introduce new
observable effects, reflected by the modified hard factors. For
example, for the dominant channel $qg\to qg$ for the dijet-correlation,
the hard factor for the single-transverse-spin
dependent cross section is about a factor $1/2$ smaller than that
for the spin-independent cross section, and the sign of this
contribution is the same as that for the Sivers asymmetry in SIDIS.
Such effects can be tested experimentally in the near
future. Some of the phenomenological consequences of this
have already been investigated in \cite{dijet-cor1}.

In our analysis, we have adopted the SIDIS definition of the TMD parton
distributions. We could also have chosen the definition
according to the Drell-Yan process. The only difference between
these two definitions is the direction of the gauge link. As is
well-known~\cite{Col02,BelJiYua02,BoeMulPij03}, the so-called naively
time-reversal-odd parton distributions change sign between these two
processes, whereas the
time-reversal-even ones remain the same. So, if we change the
definition of the TMD parton distributions to that valid for the
Drell-Yan process, the hard factors for the single transverse-spin
dependent cross section for the dijet-correlations
will change sign as well. For the spin independent
cross section, the hard factors remain the same. We stress
that the physical (hadronic) cross sections do not change with the
definition of the parton distributions.

We emphasize again that the hard factors that we have calculated
in this section are only leading order. The small transverse
momentum scale $q_\perp$ has been generated by the model TMD
distributions that we have used. We have been able to factor out
the partonic hard parts at ${\cal O}(P_\perp)$, keeping all
$q_\perp$-dependence in the TMD distributions.
In the next section, we will take a different approach and treat
the cross sections in collinear factorization, generating
$q_\perp$ perturbatively from (nearly collinear) gluon radiation.
In this way we can to some degree investigate the role of perturbative
QCD corrections, and their influence on factorization.
We will find that the generalized factorization formulas in
Eqs.~(\ref{e4}),(\ref{e2}) remain valid at this order.

\section{Collinear Factorization Approach}

In this section, we will use the collinear factorization approach to
calculate contributions to both the spin-averaged and the single
transverse-spin dependent dijet cross sections
from partonic processes with one-gluon radiation. We will work in
a kinematic region of ``intermediate'' imbalance between the two
jets: $\Lambda_{\rm QCD}\ll q_\perp \ll P_\perp$. When
$q_\perp\gg\Lambda_{\rm QCD}$, we expect perturbative QCD and the
collinear factorization approach to be valid for calculating both
the $q_\perp$- and the $P_\perp$-dependence. On the other hand, the
limit $q_\perp\ll P_\perp$ allows us to verify the generalized
factorization formulas in Eqs.~(\ref{e4}),(\ref{e2}) in the
transition region where the TMD factorization could also be valid
\cite{JiQiuVogYua06}. By an explicit calculation, we will
demonstrate that the contribution from one gluon radiation nearly
parallel to the incident hadrons can be factorized into the
same hard factors calculated in last section and the ${\cal
O}(g^2)$ perturbatively generated TMD parton distributions defined
in Eq.~(\ref{e3}), which effectively shows the validity of
the generalized TMD factorization formulas in Eqs.~(\ref{e4}) and
(\ref{e2}) at this order and in this intermediate region.

In Refs.~\cite{JiQiuVogYua06} we performed similar calculations
for the much simpler cases of the Drell-Yan and SIDIS cross sections,
with the same result. This established that the two mechanisms for
generating SSAs, one based on the Sivers function and TMD factorization,
the other on twist-3 quark-gluon correlations and collinear factorization,
are related and can in a sense be regarded as ``unified''.
It is important to point out that the consistency between the TMD
and collinear approaches in the intermediate region
$\Lambda_{\rm QCD}\ll q_\perp \ll P_\perp$ is not sufficient to
actually {\it prove} the TMD factorization formalism in the
region $q_\perp\sim \Lambda_{\rm QCD}$. But, the
consistency is certainly a necessary condition that TMD
factorization formulas like those in Eqs.~(\ref{e4}) and
(\ref{e2}) need to satisfy. It is the purpose of this section and this
paper to show this consistency for the dijet momentum imbalance
at the first non-trivial order in perturbative QCD.

We will take the simple $qq'\to qq'$ subprocess as an example to
show how this works. The extension to all other channels will
follow. For this channel the hard factors $H_{qq'\to qq'}^{uu}$
and $H_{qq'\to qq'}^{\rm Sivers}$ for the unpolarized and
polarized cross sections have been calculated in the last section.
When the dijet-imbalance is large compared to $\Lambda_{\rm QCD}$,
we can expand the factorization formulas, and the $q_\perp$
dependence will come from the TMD parton distributions at large
transverse momentum. For example, Eq.~(\ref{e2}) for the
unpolarized cross section will become,
\begin{eqnarray}
&&\frac{d^5\sigma^{uu}}{dy_1dy_2dP_\perp^2d^2\vec{q}_\perp}|_{\Lambda_{\rm
QCD}\ll q_\perp\ll P_\perp} \nonumber\\
&&~~~~~~~~~~~~~~~~~~~~~= H_{qq'\to
qq'}^{uu}\left\{x_aq(x_a,q_{\perp})x_bq(x_b)+x_aq(x_a)x_bq(x_b,q_\perp)+
\cdots\right\}
\ ,\label{e35}
\end{eqnarray}
where $H_{qq'\to qq'}^{uu}$ is the hard factor, and where we only keep
the two contributions from the incident quark distributions at
large transverse momentum. $q(x,q_\perp)$ is the TMD quark
distribution and $q(x)$ the quark distribution integrated over
transverse momentum. There could in general also be contributions
from the soft factor and/or from fragmentation functions
in the final state. In this paper, we are only interested in verifying
the part of the factorization formula that involves the TMD parton
distributions in Eq.~(\ref{e3}).

Because $q_\perp\gg \Lambda_{\rm QCD}$, the transverse-momentum-dependence
of the TMD parton distributions can be calculated in perturbative
QCD, and the results will depend on the integrated parton
distributions. This dependence is well known, see
\cite{JiQiuVogYua06} for example,
\begin{eqnarray}
q(x_a,q_\perp)&=&
\frac{\alpha_s}{2\pi^2}\frac{1}{\vec{q}_\perp^2}C_F\int\frac{dx}{x}
q(x) \left[\frac{1+\xi^2}{1-\xi}+\cdots\right]\ , \label{e36}
\end{eqnarray}
where $C_F=(N_c^2-1)/2N_c$, and $q(x)$ is the integrated quark
distribution as mentioned above and $\xi=x_a/x$. Note that we only
keep the collinear part in the distribution, i.e., $1-\xi\neq 0$.
It will certainly be necessary to consider the soft contribution
at $\xi=1$~\cite{JiQiuVogYua06}. Similarly, we can calculate
$q(x_b,q_\perp)$ at large $q_\perp$. Substituting these two
results into Eq.~(\ref{e35}), we obtain
\begin{eqnarray}
\frac{d^5\sigma^{uu}}{dy_1dy_2dP_\perp^2d^2\vec{q}_\perp}|_{\Lambda_{\rm
QCD}\ll q_\perp\ll P_\perp} &=& H_{qq'\to
qq'}^{uu}\frac{1}{\vec{q}_\perp^2}\frac{\alpha_s}{2\pi^2}
C_F\int{dxdx'}q(x)q(x')\nonumber\\
&&\times {\xi \xi'}
\left\{\frac{1+\xi^2}{(1-\xi)}\delta(\xi'-1)+\frac{1+\xi^{\prime2}}
{(1-\xi')}\delta(\xi-1)\right\}
\ , \label{e37}
\end{eqnarray}
where $\xi$ is as defined above and $\xi'=x_b/x'$. In the next subsection,
we will demonstrate by explicit calculation that the above result
is reproduced by the direct calculation in the collinear
factorization approach.

Similarly, the Sivers function $q_T(x_a,k_\perp)$ at large
$k_\perp$ can be calculated in perturbative QCD, and the result
depends on the twist-3 quark-gluon correlation function $T_F$ (the
Qiu-Sterman matrix element)~\cite{qiusterman,twist3-new}:
\begin{eqnarray}
T_F(x_1,x_2) &\equiv & \int\frac{d\zeta^-d\eta^-}{4\pi} e^{i(x_1
P_A^+\eta^-+(x_2-x_1)P_A^+\zeta^-)}
\nonumber \\
&\times & \epsilon_\perp^{\beta\alpha}S_{\perp\beta} \,
\left\langle P_A,S|\overline\psi(0){\cal L}(0,\zeta^-)\gamma^+
\right. \label{TF}\\
&\times & \left. g{F_\alpha}^+ (\zeta^-) {\cal L}(\zeta^-,\eta^-)
\psi(\eta^-)|P_A,S\right\rangle  \ , \nonumber
\end{eqnarray}
where ${\cal L}$ is the proper gauge link to make the matrix
element gauge invariant, and where the sums over color and spin
indices are implicit. Keeping again only the contribution from the
collinear part, we will have \cite{JiQiuVogYua06},
\begin{eqnarray}
q_T^{\rm SIDIS}(x_a,k_\perp)&=&-\frac{\alpha_s}{4\pi^2}\frac{2M_P}
{(\vec{k}_\perp^2)^2}\int\frac{dx}{x} \left\{\frac{1}{2N_C}
\left\{ \left[x\frac{\partial}{\partial x}T_F(x,x)\right](1+\xi^2)
      +T_F(x,x-\widehat{x}_g)\frac{1+\xi}{(1-\xi)}\nonumber\right.\right.\\
 &&\left. \left.    +T_F(x,x)\frac{(1-\xi)^2(2\xi+1)-2}{(1-\xi)}\right\}
+C_F T_F(x,x-\widehat{x}_g)\frac{1+\xi}{(1-\xi)_+} \right\} \ ,
\label{e38}
\end{eqnarray}
where as above $\xi=x_a/x$ and $\widehat x_g=(1-\xi)x$. The definition
of the quark Sivers function in Eq.~(\ref{e3}) has been used to
obtain the above result. We emphasize again that in this
definition the gauge link is simple in the Feynman gauge, and goes
to $+\infty$. As mentioned above, we have to take into account the
perturbative expansion of the gauge link up to second order
(${\cal O}(g^2)$, see also the diagrams drawn in
\cite{JiQiuVogYua06}). Substituting the above result and the
unpolarized quark distribution $q(x_b,k_\perp)$ of
Eq.~(\ref{e36}) into the factorization formula Eq.~(\ref{e4}), we
obtain the single-spin dependent cross section in the
intermediate transverse momentum region as
\begin{eqnarray}
\frac{d^5\Delta\sigma(S_\perp)}
     {dy_1dy_2dP_\perp^2d^2 \vec{q}_\perp}
     |_{\Lambda_{\rm QCD}\ll q_\perp\ll P_\perp}
&=& -H_{qq'\to qq'}^{\rm Sivers}
\frac{\epsilon^{\alpha\beta}S_\perp^\alpha q_\perp^\beta}
     {(\vec{q}_\perp^2)^2}
\frac{\alpha_s}{2\pi^2} \left\{x_bq(x_b)\int dx \xi
     \left[\frac{1}{2N_c}(1+\xi^2)
\nonumber\right.\right.\\
&&\times \left(x\frac{\partial}{\partial x}T_F(x,x)\right)
+\frac{1}{2N_c}T_F(x,x)\frac{2\xi^3-3\xi^2-1}{1-\xi}
\nonumber\\
&&\left. +\left(\frac{1}{2N_c}+C_F\right) T_F(x,x-\hat x_g)
 \frac{1+\xi}{1-\xi}\right]
\nonumber\\
&&\left. + x_a T_F(x_a,x_a)C_F \int d x' \xi'
 \frac{1+\xi^{\prime 2}}{1-\xi'}
\right\} \ . \label{e39}
\end{eqnarray}

In the above result, the term in square brackets comes from the
Sivers function in the polarized nucleon at large-$k_\perp$,
while the last term corresponds to the large-$k_\perp$ unpolarized
quark distribution from the unpolarized nucleon. We have also used
the leading order relation
\cite{BoeMulPij03}
\begin{equation}
\frac{1}{M_P}\int d^2k_{\perp}\, \vec{k}^2_\perp\, q_T^{\rm
SIDIS}(x,k_\perp) = - T_F(x,x)\, , \label{e34}
\end{equation}
to derive the second term.

\begin{figure}[t]
\begin{center}
\includegraphics[width=11cm]{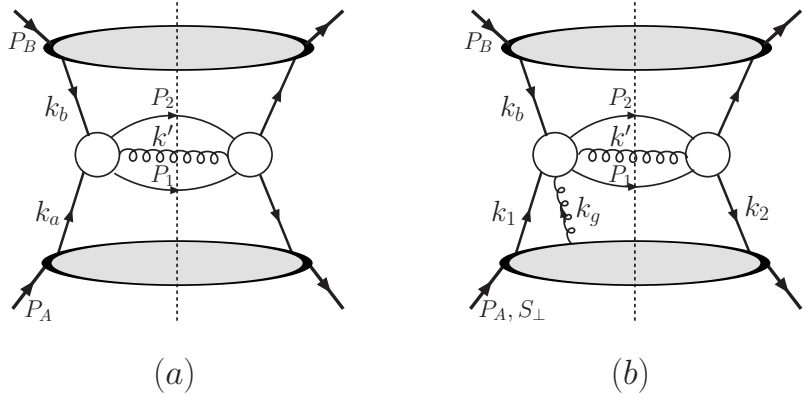}
\end{center}
\vskip -0.4cm \caption{\it Generic diagrams for quark-quark scattering
contributions to the unpolarized and the single transverse-spin dependent
cross sections in the collinear factorization approach.}
\label{generic_diagrams}
\end{figure}

In the following, we will derive the above spin-averaged and
spin-dependent differential cross sections in the collinear
factorization framework in the regime $P_\perp \gg q_\perp \gg
\Lambda_{\rm QCD}$, and verify the expressions in
Eqs.~(\ref{e37}),(\ref{e39}).  This will provide a clear
demonstration of the consistency between the collinear
factorization approach and the TMD factorization formalism in the
region of intermediate dijet imbalance. In
Fig.~\ref{generic_diagrams}, we draw the generic Feynman diagrams
for the unpolarized and single-transverse-spin dependent cross
sections for the dijet-correlation in the collinear factorization
approach. In these diagrams, $P_A$ and $P_B$ are the momenta of
the two incident hadrons. For the single-transverse polarized
scattering, $P_A$ labels the polarized hadron. $P_1$ and $P_2$ are
the momenta of the two jets in the final states. $k_a$ and $k_b$
are the momenta of the two incoming partons, and for polarized
scattering $k_{1}$ and $k_{2}$ will be needed
to define the complete kinematics for the momentum flow associated
with the polarized hadron. The blob in the center represents
tree-level Feynman diagrams with the given initial- and
final-state partons. In the collinear approach, jets are produced
back-to-back by the Born diagrams, so that there will be no
imbalance between the two jets if there is no additional gluon
radiation. In order to obtain an imbalance, we have to add
at least one gluon (momentum $k'$) into the final state, as shown in
Fig.~\ref{generic_diagrams}. The transverse momentum of $k'$ will be
equal to the imbalance of the two jets, i.e.,
$\vec{k}'_\perp=-\vec{q}_\perp=-(\vec{P}_{1\perp}+\vec{P}_{2\perp})$.

We will use the simple $qq'\to qq'g$ subprocess as an example to
demonstrate the method and details of the derivation.
All other partonic channels can be analyzed in a similar
manner. In subsection A, we will perform the calculation for
unpolarized cross section, and in subsection B, we will deal with
the single transverse-spin dependent cross section.

\subsection{Unpolarized $q q'\to q q' g$}

In terms of collinear QCD factorization for unpolarized hadronic
collisions \cite{ColSopSte89}, the contribution by the $qq'\rightarrow
qq'g$ subprocess to the dijet cross section can be written as (see
Fig.~\ref{generic_diagrams}(a)):
\begin{equation}
\frac{d\, \sigma_{(qq')}^{uu}}
     {dy_1dy_2dP_\perp^2d^2\vec{q}_\perp}
=\int \frac{dx}{x}\, \frac{dx'}{x'}\, q(x)\, q'(x')\,
\frac{1}{16\, s\, (2\pi)^4}|\overline{\cal M}_{qq'\to qq'g}|^2\,
\delta((k')^2) \, ,
\label{x-avg}
\end{equation}
where $s=(P_A+P_B)^2$, $|\overline{\cal M}_{qq'\to qq'g}|^2$ is
the spin and color averaged scattering amplitude squared
for the partonic subprocess $qq'\to qq'g$,
and $q(x)$ and $q'(x')$ are quark distributions for the
incoming hadrons, $A$ and $B$, respectively, at momentum fractions
$x$ and $x'$. In the collinear factorization approach, initial-state quark
momenta are approximated as $k_a^\mu = x P_A^\mu$ and
$k_b^\mu = x' P_B^\mu$. Neglecting the invariant mass of the jet
which is much smaller than the jet's energy,
and keeping only leading powers in $q_\perp/P_\perp$,
we can write the jet momenta as
$P_1 = P_\perp\, (\frac{e^{y_1}}{\sqrt{2}},
                  \frac{e^{-y_1}}{\sqrt{2}}, 1)$
and $P_2 = P_\perp\, (\frac{e^{y_2}}{\sqrt{2}},
                      \frac{e^{-y_2}}{\sqrt{2}}, -1)$.
In Eq.~(\ref{x-avg}) and the rest of this paper, the dependence on
factorization and renormalization scales is suppressed.

\begin{figure}[t]
\begin{center}
\includegraphics[width=13cm]{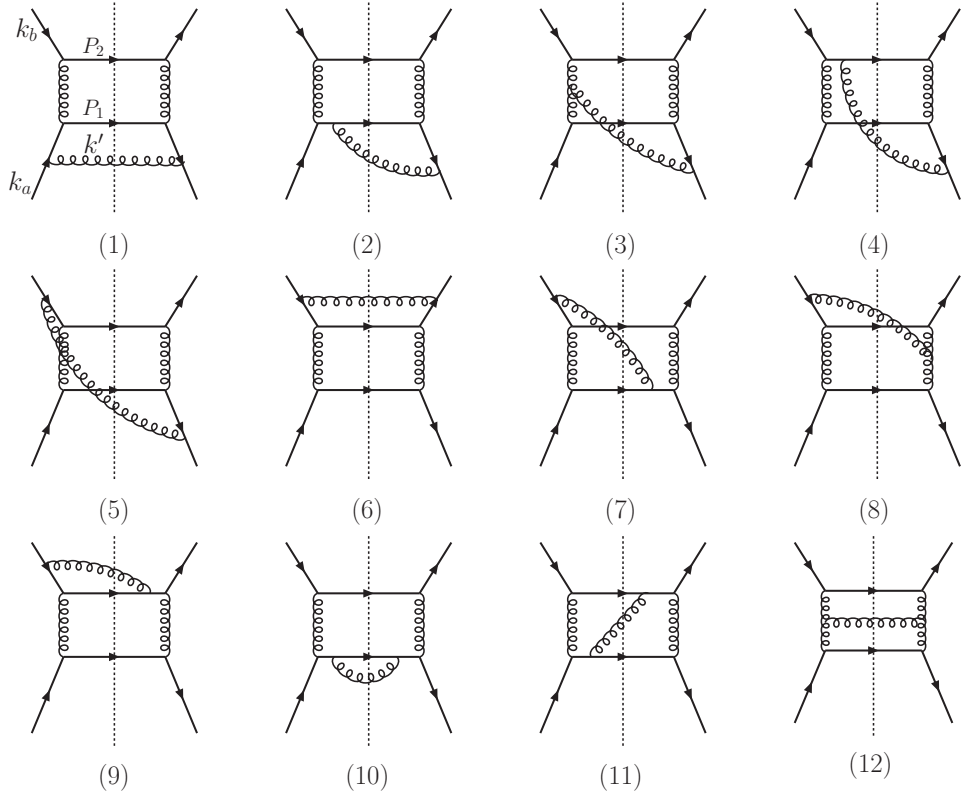}
\end{center}
\vskip -0.4cm \caption{\it Some of the Feynman diagrams
for the $qq'\to qq'$ channel contributing to the dijet imbalance
in the collinear factorization approach. Diagrams (1-5)
contribute to the cross section at leading power when $k'$ is
parallel to $P_A$, which can be attributed to the parton
distribution in hadron $A$, whereas (5-9) contribute when $k'$ is
parallel to $P_B$ and belong to the parton distribution in hadron $B$.
All other diagrams are power suppressed when $k'$ is parallel to
either $P_A$ or $P_B$ and will contribute to other factors
in the factorization formula.}\label{f13}
\end{figure}

The basic $qq'\to qq'$ process only has one Feynman
diagram. Radiation of an additional gluon leads to a drastic increase in
the number of diagrams.  In Fig.~\ref{f13}, we show sample diagrams
that are most relevant for our analysis in the limit $P_\perp \gg
q_\perp \gg \Lambda_{\rm QCD}$. We work in Feynman gauge, and
classify these diagrams into different groups by using the power
counting technique for this limit \cite{ColSopSte89}. For example,
if the momentum $k'$ of the radiated gluon is nearly parallel to $P_A$,
only diagrams (1-5) contribute at leading power in $q_\perp/P_\perp$, and
all other diagrams are suppressed by $q_\perp/P_\perp$. The contributions
by these five diagrams will be factorized into the TMD quark
distribution of the hadron with momentum $P_A$. Similarly, if $k'$ is
nearly parallel to $P_B$, only (5-9) contribute at leading power, and
their contributions can be factorized into the TMD quark
distribution of the hadron with momentum $P_B$. Diagrams (10,11) and
other similar diagrams will contribute at leading power when
$k'$ is parallel to $P_1$, the momentum of one of the outgoing
partons. Such contributions will become part of the jet and
be subject to the jet definition. If instead of the jet a
hadron is observed, these contributions belong to a TMD fragmentation
function. Diagram (12) does not contribute to any of the above cases,
but will contribute when $k'$ becomes soft and thus give rise to part of
the soft factor in
the factorization formula. As mentioned above, in this paper we will
study the factorization of the collinear gluon interactions into
the gauge invariant TMD parton distributions. We will therefore focus
on the contributions when $k'$ is parallel to either $P_A$ or $P_B$. The
other contributions will be important to fully understand the
factorization formalism, but we leave them to a future publication.

In the following, we will work out the result for diagrams (1-5)
in some detail, in order to see how their contributions can be
factorized into the TMD quark distribution of hadron $A$.
We choose the kinematics so that the momenta of two incident hadrons
only have light-cone ``plus'' and ``minus''
components, respectively: $P_A=(P_A^+,0^-,0_\perp)$ and
$P_B=(0^+,P_B^-,0_\perp)$. Then, in the collinear approximation,
the momentum $k_a$ of the initial-state parton $a$ only has a plus
component while $k_b$ has only a minus component.
For extracting the leading contribution in $q_\perp/P_\perp$,
it is convenient to decompose the total dijet momentum
$q=P_1+P_2$ and the momentum $k'$ of the ``unobserved'' final-state gluon
into
\begin{eqnarray}
q&=& \xi\, k_a + \xi'\, k_b + q_\perp\ ,
\nonumber \\
k'&=&(1-\xi)\, k_a+(1-\xi')\, k_b +k'_\perp\ ,
\label{kprime}
\end{eqnarray}
where $\vec{k}_\perp'=-\vec{q}_\perp$, and $\xi$ and $\xi'$ are
the fractions of the initial quark momenta carried into the
dijet final state. The partonic Mandelstam variables
are given at leading power in $q_\perp/P_\perp$ by
\begin{eqnarray}
\hat s&=& (\xi k_a + \xi' k_b)^2
       = \xi\xi' 2k_a\cdot k_b = 2P_1\cdot P_2
       \ ,\nonumber\\
\hat t&=& (\xi k_a - P_1)^2
       = -\xi 2k_a\cdot P_1 =-\xi' 2k_b\cdot P_2
       \ ,\nonumber\\
\hat u&=& (\xi k_a - P_2)^2
       = -\xi 2k_a\cdot P_2 =-\xi' 2k_b\cdot P_1\, ,
\label{e44}
\end{eqnarray}
where $\xi=x_a/x$ and $\xi'=x_b/x'$ with
$x_a=\frac{P_\perp}{\sqrt{s}}\, (e^{y_1}+e^{y_2})$,
$x_b=\frac{P_\perp}{\sqrt{s}}\, (e^{-y_1}+e^{-y_2})$, and
$\hat{s}=x_a x_b s$.

\begin{figure}[t]
\begin{center}
\includegraphics[width=13cm]{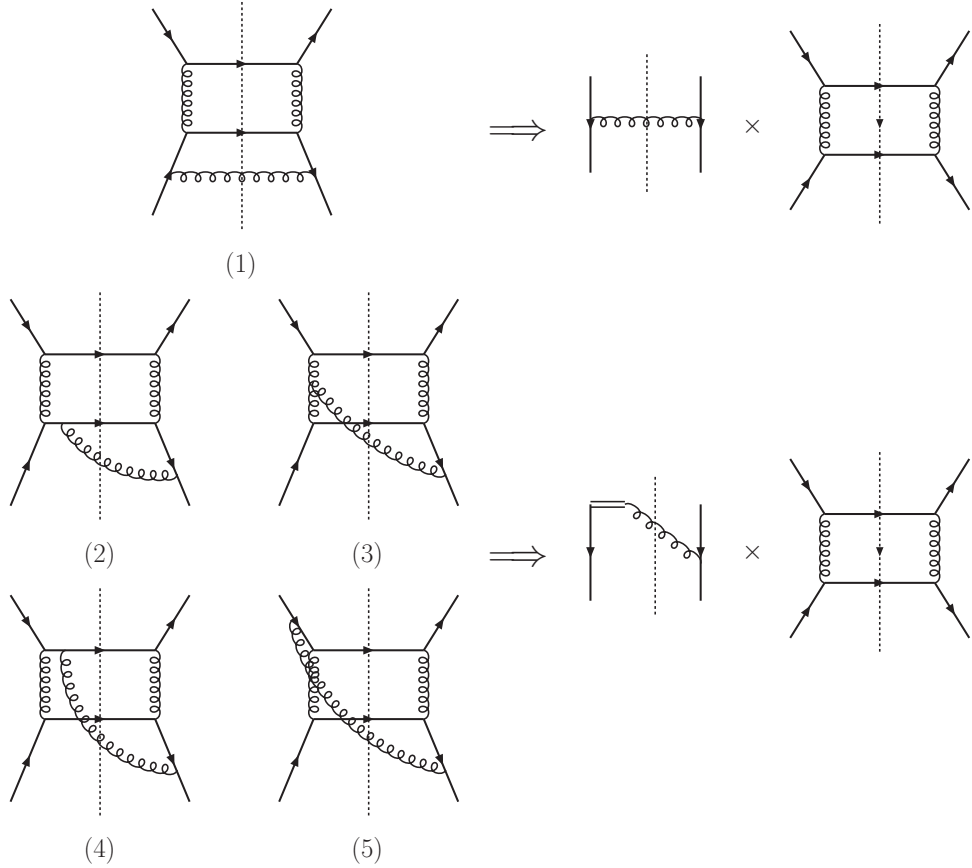}
\end{center}
\vskip -0.4cm \caption{\it Factorization of the contributions from
diagrams (1-5) into the TMD quark distribution of hadron $A$ times the
hard factor, when the momentum $k'$ of the radiated gluon is nearly
parallel to $P_A$. These five diagrams are the leading ones among those in
Fig.~\ref{f13}.} \label{f14}
\end{figure}

When $k'$ is parallel to $P_A$ and $k'_\perp\ll P_\perp$,
the on-shell condition
$k'^2=(1-\xi)(1-\xi')(2k_a\cdot k_b) - (\vec{k}'_\perp)^2=0$
leads to $(1-\xi')\approx 0$ while $(1-\xi)$ remains large.
The delta function in Eq.~(\ref{x-avg}) can then be reduced to
\begin{equation}
\delta \left((k')^2\right)\longrightarrow
\frac{\xi}{\hat s}\frac{\delta(1-\xi')}{1-\xi} \, .
\label{e45}
\end{equation}
With this approximation, the squared amplitude will become much
simpler. For example, the contribution by diagram (1) will be
\begin{equation}
|\overline{{\cal
M}}|^2_{(1)}=\frac{4g^6}{\xi}\frac{(1-\xi)^2}{\vec{q}_\perp^2}\frac{\hat
s^2+\hat u^2}{\hat t^2}\ ,
\end{equation}
where the color-factor has not been included. For this diagram,
it is very simple, and can be easily factorized as
\begin{equation}
\frac{1}{N_c^2}{\rm Tr}(T^aT^aT^bT^c){\rm Tr}(T^bT^c)=C_F\times
\frac{1}{N_c^2}{\rm Tr}(T^bT^c){\rm Tr}(T^bT^c)=C_F\times C_u\ ,
\end{equation}
where $C_u=(N_c^2-1)/4N_c^2$ is the color-factor for the partonic
process $qq'\to qq'$. Substituting the above results into
Eq.~(\ref{x-avg}), we obtain the contribution to the differential
cross section from this diagram as
\begin{equation}
\frac{d\sigma_{(qq')}^{uu}}
     {dy_1dy_2dP_\perp^2d^2\vec{q}_\perp}|_{{\mathrm{Fig.}}\ref{f13}(1)}
= \left[\frac{\alpha_s^2\pi}{\hat s^2}C_u
      \frac{2(\hat s^2+\hat u^2)}{\hat t^2}
\right] x_bq(x_b) \int dx\, q(x) \frac{\alpha_s}{2\pi^2}C_F
\frac{1}{\vec{q}_\perp^2}\xi(1-\xi) \, ,
\end{equation}
where the factor in the square brackets is the hard factor
$H_{qq'\to qq'}^{uu}$ for the TMD factorization formula,
Eq.~(\ref{e3}), and is the same as that calculated in the last
section in Eq.~(\ref{hqqp}). Other than $x_b q(x_b)$, the rest of
the right-hand-side of the above equation is a part of the TMD
quark distribution in Eq.~(\ref{e36}). We illustrate this
factorization in the upper panel of Fig.~\ref{f14}.

Diagrams (2-5) are more complicated, but their calculations are
straightforward using the above techniques. First, their color
factors can be formulated as follows:
\begin{eqnarray}
{\rm (2)}&:& \frac{1}{N_c^2}{\rm Tr}(T^aT^bT^aT^c){\rm
Tr}(T^bT^c)=C_F\times C_u+\frac{1}{N_c^2}\frac{i}{2}f_{abc}{\rm
Tr}(T^aT^bT^c)
\nonumber\ , \\
{\rm (3)}&:& \frac{1}{N_c^2}(-if_{adc}){\rm Tr}(T^aT^bT^c){\rm
Tr}(T^bT^d)=-\frac{1}{N_c^2}\frac{i}{2}f_{abc}{\rm Tr}(T^aT^bT^c)
\nonumber\ , \\
{\rm (4)}&:& \frac{1}{N_c^2}{\rm Tr}(T^aT^bT^c){\rm
Tr}(T^aT^cT^b) \nonumber \ ,\\
{\rm (5)}&:& \frac{1}{N_c^2}{\rm Tr}(T^aT^bT^c){\rm Tr}(T^aT^bT^c)
\ , \label{e49}
\end{eqnarray}
where we notice that the only difference between the color factors
for diagrams (4) and (5) is the $T^c T^b$ vs. $T^b T^c$ in the
second trace. After applying
Eqs.~(\ref{e44}),(\ref{e45}), the squared amplitudes (without the
color-factors) for these diagrams are reduced to the following
simple expressions:
\begin{eqnarray}
|\overline{{\cal M}}|^2_{(2)}&=&4g^6\frac{\hat s^2+\hat u^2}{\hat
t^2}\frac{1}{\vec{q}_\perp^2} \ , \nonumber\\
|\overline{{\cal M}}|^2_{(3)}&=&4g^6\frac{\hat s^2+\hat u^2}{\hat
t^2}\frac{1-\xi}{\vec{q}_\perp^2} \ , \nonumber\\
|\overline{{\cal M}}|^2_{(4)}&=&4g^6\frac{\hat s^2+\hat u^2}{\hat
t^2}\frac{\xi}{\vec{q}_\perp^2} \ , \nonumber\\
|\overline{{\cal M}}|^2_{(5)}&=&-|\overline{{\cal M}_{(4)}}|^2\ ,
\end{eqnarray}
where the contributions from diagrams (4) and (5) are equal
and opposite.  Thus, when color is included, their sum will
depend on the difference of the color-factors in Eq.~(\ref{e49}),
\begin{equation}
\frac{1}{N_c^2}{\rm Tr}(T^aT^bT^c){\rm
Tr}(T^aT^cT^b) -
\frac{1}{N_c^2}{\rm Tr}(T^aT^bT^c){\rm Tr}(T^aT^bT^c)=
-\frac{1}{N_c^2}\frac{i}{2}f_{abc}{\rm Tr}(T^aT^bT^c) \ .
\end{equation}
From the above results, we find that the contributions from diagrams
(2-5) only have two color-factors: $C_F\times C_u$ and $-if_{abc}{\rm
Tr}(T^aT^bT^c)$. However, it is straightforward to see that the
contributions coming with the color structure $-if_{abc}{\rm
Tr}(T^aT^bT^c)$ cancel each other, so that the only
contribution will have the color factor $C_F\times C_u$.
Therefore, the final result for the amplitude squared from diagrams (2-5)
will be, including the color-factor,
\begin{equation}
|\overline{\cal M}|^2_{(2-5)}=C_F\times C_u \frac{4g^6(\hat
s^2+\hat u^2)}{\hat t^2}\frac{1}{\vec{q}_\perp^2} \ .
\end{equation}
Inserting the above into Eq.~(\ref{x-avg}), we find the
contributions by diagrams (2-5) to the differential cross section:
\begin{equation}
\frac{d\sigma_{(qq')}^{uu}}
     {dy_1dy_2dP_\perp^2d^2\vec{q}_\perp}|_{{\mathrm{Fig.}}\ref{f13}(2-5)}
=H_{qq'\to qq'}^{uu}x_bq(x_b)\int dxq(x)
\frac{\alpha_s}{2\pi^2}C_F\frac{1}{\vec{q}_\perp^2}\frac{\xi^2}{1-\xi}
\ ,
\end{equation}
where $H_{qq'\to qq'}^{uu}$ is the hard factor for the TMD
factorization formula mentioned above. Clearly, apart from $x_b
q(x_b)$, this result can also be interpreted as part of the TMD
quark distribution for hadron $A$, multiplied by the hard factor.
We display this factorization in the lower panel of
Fig.~\ref{f14}. It is important to note that we have taken into
account all possible attachments of the radiated gluon to the hard
part to obtain the right answer. Especially, the diagram with the
gluon attached to the gluon propagator is not power suppressed,
and its contribution is crucial to achieve the final
factorization. Summation over all gluon attachments leads to a
gauge link contribution as shown in Fig.~\ref{f14}. Of course,
this result is also consistent with the Ward identity that allows
to factorize all gluon attachments into a single gauge link.

After summing all the contributions by diagrams (1-5),
including the mirror diagrams of (2-5), we obtain the final result for
the differential cross section for the dijet-correlation when the
radiated gluon momentum $k'$ is nearly parallel to $P_A$:
\begin{equation}
\left. \frac{d\sigma_{(qq')}^{uu}}
     {dy_1dy_2dP_\perp^2d^2\vec{q}_\perp}
\right|_{k'\propto P_A} = H_{qq'\to
qq'}^{uu}(\hat{s},\hat{t},\hat{u})\,
  \left[ x_b\, q'(x_b)\right]
  \int dx\, q(x) \left[
  \frac{\alpha_s}{2\pi^2}\,C_F\,
  \frac{1}{\vec{q}_\perp^2}\,\xi\frac{1+\xi^2}{1-\xi} \right]
\, ,
\label{x-avg-pa}
\end{equation}
which reproduces the first term in the TMD expansion result in
Eq.~(\ref{e37}).

Similarly, from diagrams (5-9) and the corresponding mirror
diagrams, we evaluate the leading contribution in
$q_\perp/P_\perp$ when the gluon momentum $k'$ is nearly parallel to
$P_B$. Combining with Eq.~(\ref{x-avg-pa}), we find:
\begin{eqnarray}
\left. \frac{d\sigma_{(qq')}^{uu}}
     {dy_1dy_2dP_\perp^2d^2\vec{q}_\perp}
\right|_{P_\perp\gg q_\perp\gg \Lambda_{\rm QCD}}^{{\mathrm{coll.~fac.}}} &=&
H_{qq'\to qq'}^{uu}\, \frac{1}{\vec{q}_\perp^2}\,
\frac{\alpha_s}{2\pi^2}\, C_F
\int dx\, dx'\, q(x)\, q'(x') \nonumber\\
&\times &
{\xi\, \xi'}
\left[\frac{1+\xi^2}{(1-\xi)}\delta(\xi'-1)
+\frac{1+\xi^{\prime2}}{(1-\xi')}\delta(\xi-1)\right]
\label{x-avg-qq}
\end{eqnarray}
with $\xi=x_a/x$ and $\xi'=x_b/x'$ as introduced above. We notice
that in Eq.~(\ref{x-avg-qq}) the physics at the scale $q_\perp$ is
decoupled from that at $P_\perp$, reproducing the first
order TMD factorization result in Eq.~(\ref{e37}).

In summary, by explicit calculations of the Feynman diagrams for
the partonic process $qq'\to qq'g$ which contributes to the
dijet-correlations in hadronic reactions, we have demonstrated that
when the momentum of the radiated gluon $k'$
is almost parallel to either $P_A$ or $P_B$, the contribution to
the differential cross section in Eq.~(\ref{x-avg-qq}) can be
factorized into the perturbatively generated TMD quark
distributions defined in Eq.~(\ref{e3}) for the two incident
hadrons, multiplied by the same hard factor at ${\cal O}(P_\perp)$
as calculated in the last section using the lowest-order $2\to 2$
Born diagrams.  This can be viewed as support for the generalized
factorization formalism in Eq.~(\ref{e2}) for the dijet-correlations at
hadron colliders.

\subsection{Single transverse-spin dependent cross section}

In this section we calculate the leading contribution to single
transverse-spin dependent cross section $\Delta\sigma(S_\perp)$ in
the limit $P_\perp \gg q_\perp \gg \Lambda_{\rm QCD}$ in the
twist-3 ETQS approach~\cite{et,qiusterman}. The difference between
the unpolarized cross section calculated in the last subsection
and the transverse-spin dependent one is that the latter involves
an additional
gluon from the polarized proton, which
interacts with partons in the hard part, as shown in
Fig.~\ref{generic_diagrams}(b). When both $P_\perp$ and $q_\perp$
are much larger than the typical hadronic scale $\Lambda_{\rm QCD}$,
the collinear factorization approach should be valid for
describing the SSA \cite{qs-ssa-fac},
and a nonvanishing $\Delta\sigma(S_\perp)$ is generated by the
ETQS mechanism in terms of twist-three transverse-spin dependent
quark-gluon correlation functions \cite{et,qiusterman}.  In the
ETQS formalism, the contribution of the subprocess $(g)qq'\to
qq'g$ to $\Delta\sigma(S_\perp)$, shown in
Fig.~\ref{generic_diagrams}(b), is generically given by
\begin{equation}
\frac{d\Delta \sigma(S_\perp)_{(qq')}}
     {dy_1dy_2dP_\perp^2d^2\vec{q}_\perp}
=\int \frac{dx'}{x'}\, {dx_1dx_2}\,
      T_F(x_1,x_2)\, q'(x')\,
      \frac{1}{16s(2\pi)^4}\,\delta((k')^2)\,
      {\cal H}_{(g)qq'\to qq'g}\ ,
\label{x-ssa}
\end{equation}
where ${\cal H}$ represents a partonic hard part, $x_1$ and $x_2$
are the momentum fractions of the quarks from the polarized hadron
$A$ on the two sides of the cut shown in
Fig.~\ref{generic_diagrams}(b), and $T_F(x_1,x_2)$ is the
corresponding twist-three quark-gluon correlation function defined
in Eq.~(\ref{TF}), extracted from the lower blob in the figure
\cite{qiusterman,twist3-new}.

\begin{figure}[t]
\begin{center}
\includegraphics[width=15cm]{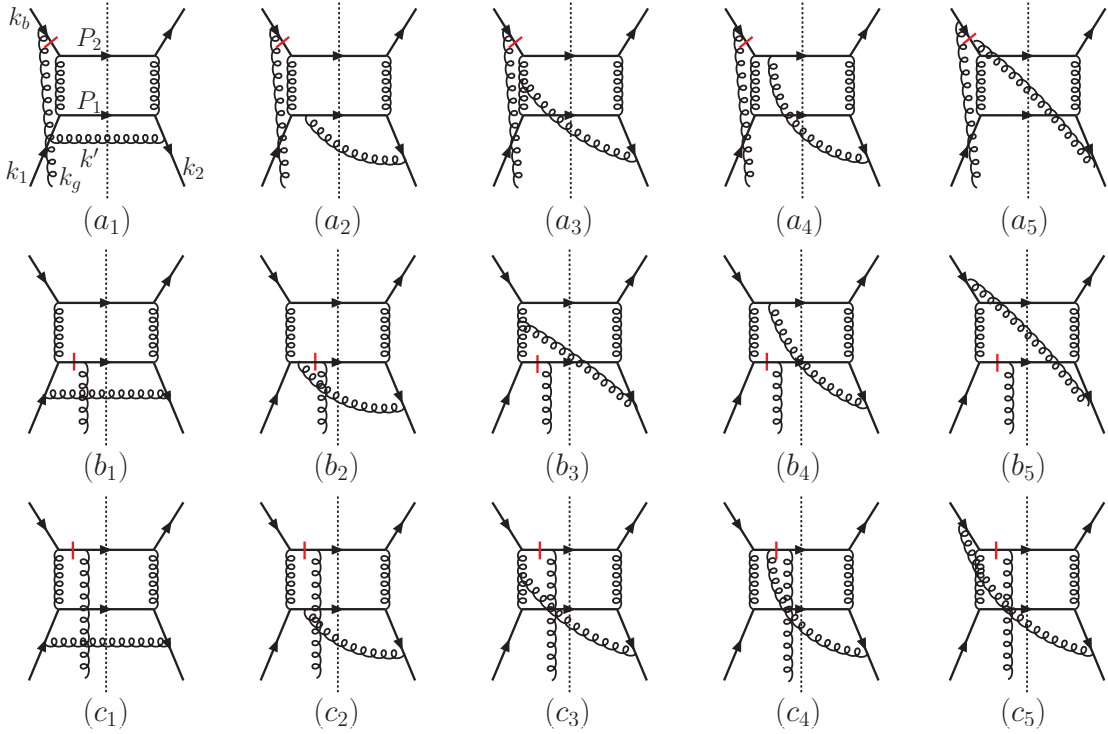}
\end{center}
\vskip -0.4cm \caption{\it Soft pole contributions when $k'$ is
parallel to $P_A$. The mirror diagrams where the polarized gluon
attaches to the right hand of the cut line are not shown but
included in the final results.}\label{f15}
\end{figure}

\begin{figure}[t]
\begin{center}
\includegraphics[width=15cm]{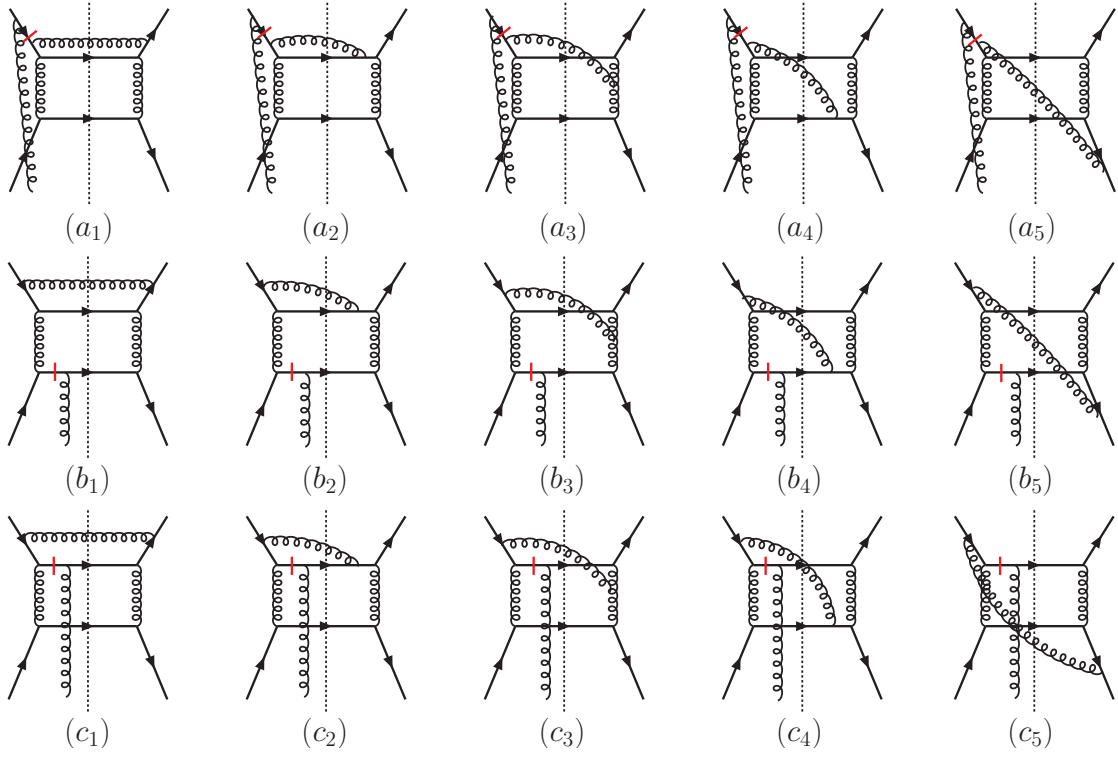}
\end{center}
\vskip -0.4cm \caption{\it Soft pole contributions when $k'$ is
parallel to $P_B$.}\label{f20}
\end{figure}

\begin{figure}[t]
\begin{center}
\includegraphics[width=10cm]{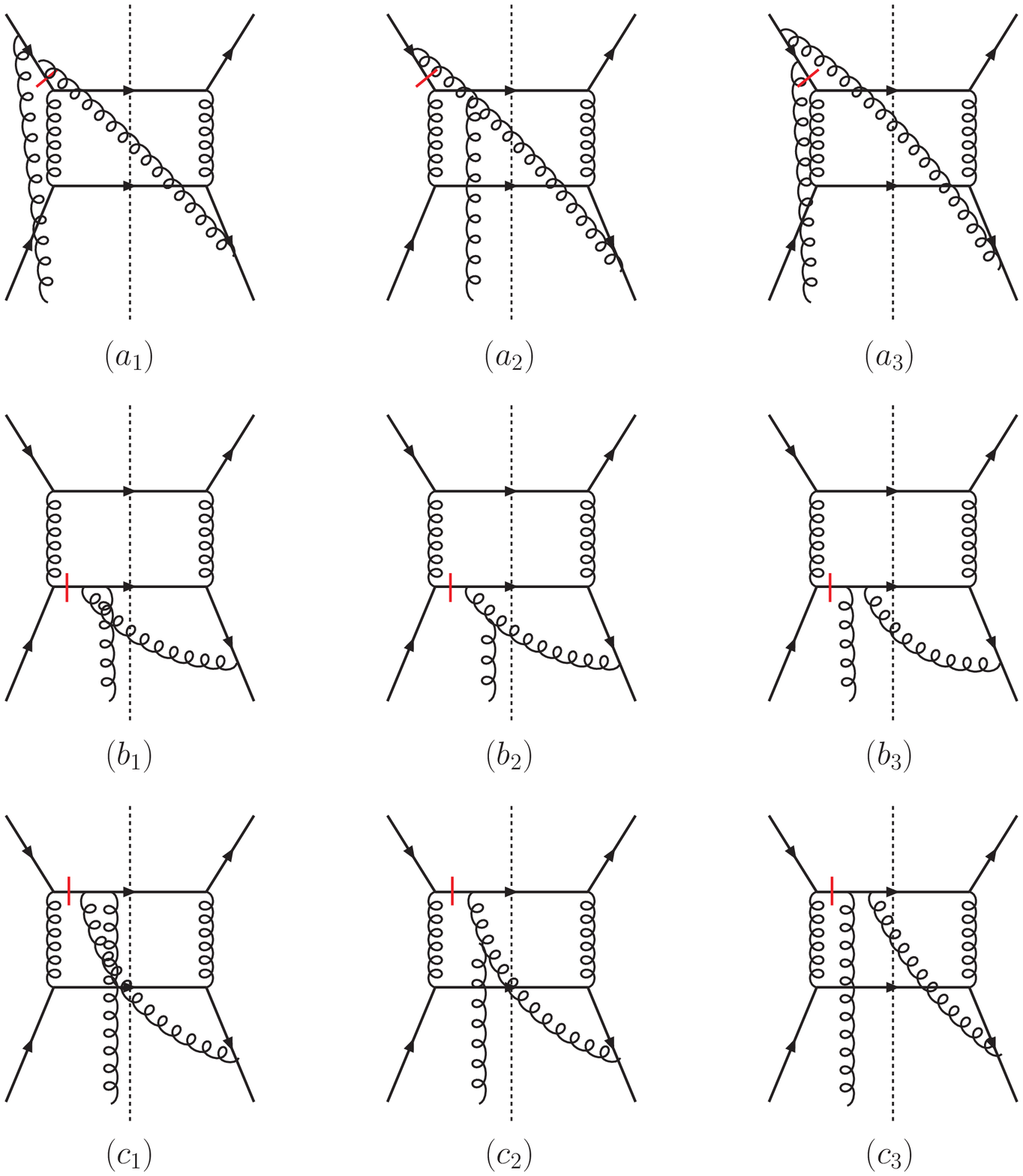}
\end{center}
\vskip -0.4cm \caption{\it Hard pole contributions when $k'$ is
parallel to $P_A$.}\label{f18}
\end{figure}

\begin{figure}[]
\begin{center}
\includegraphics[width=15cm]{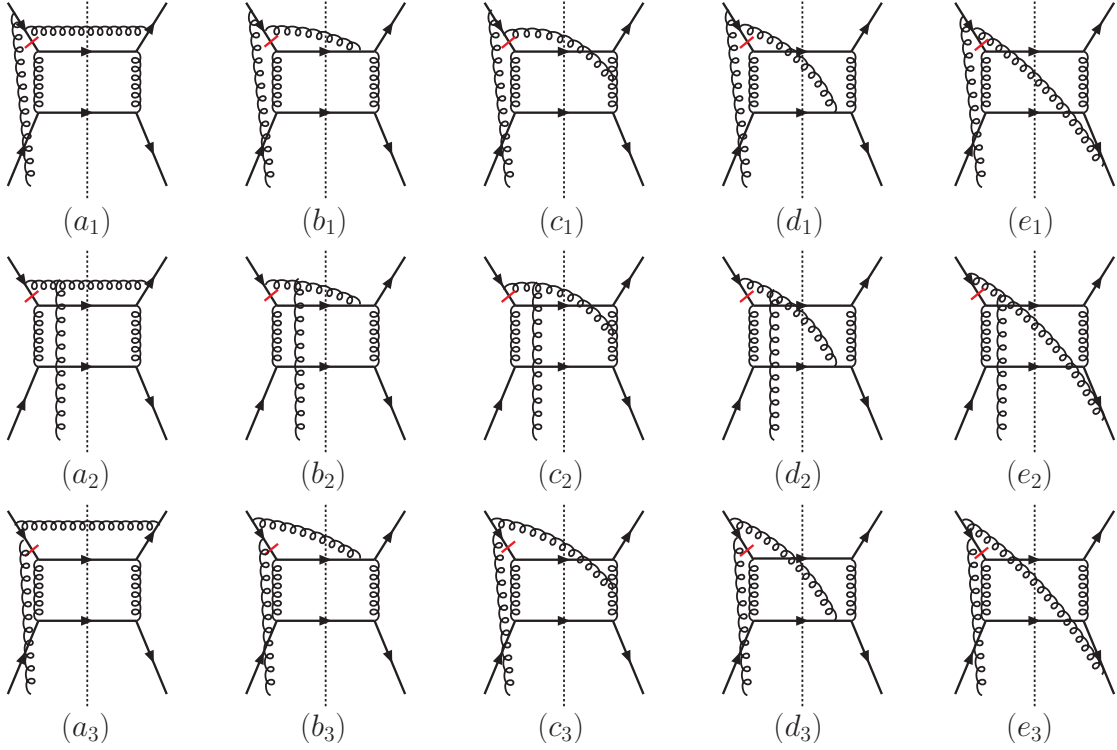}
\end{center}
\vskip -0.4cm \caption{\it Hard pole contributions when $k'$ is
parallel to $P_B$.}\label{f21}
\end{figure}

\begin{figure}[]
\begin{center}
\includegraphics[width=10cm]{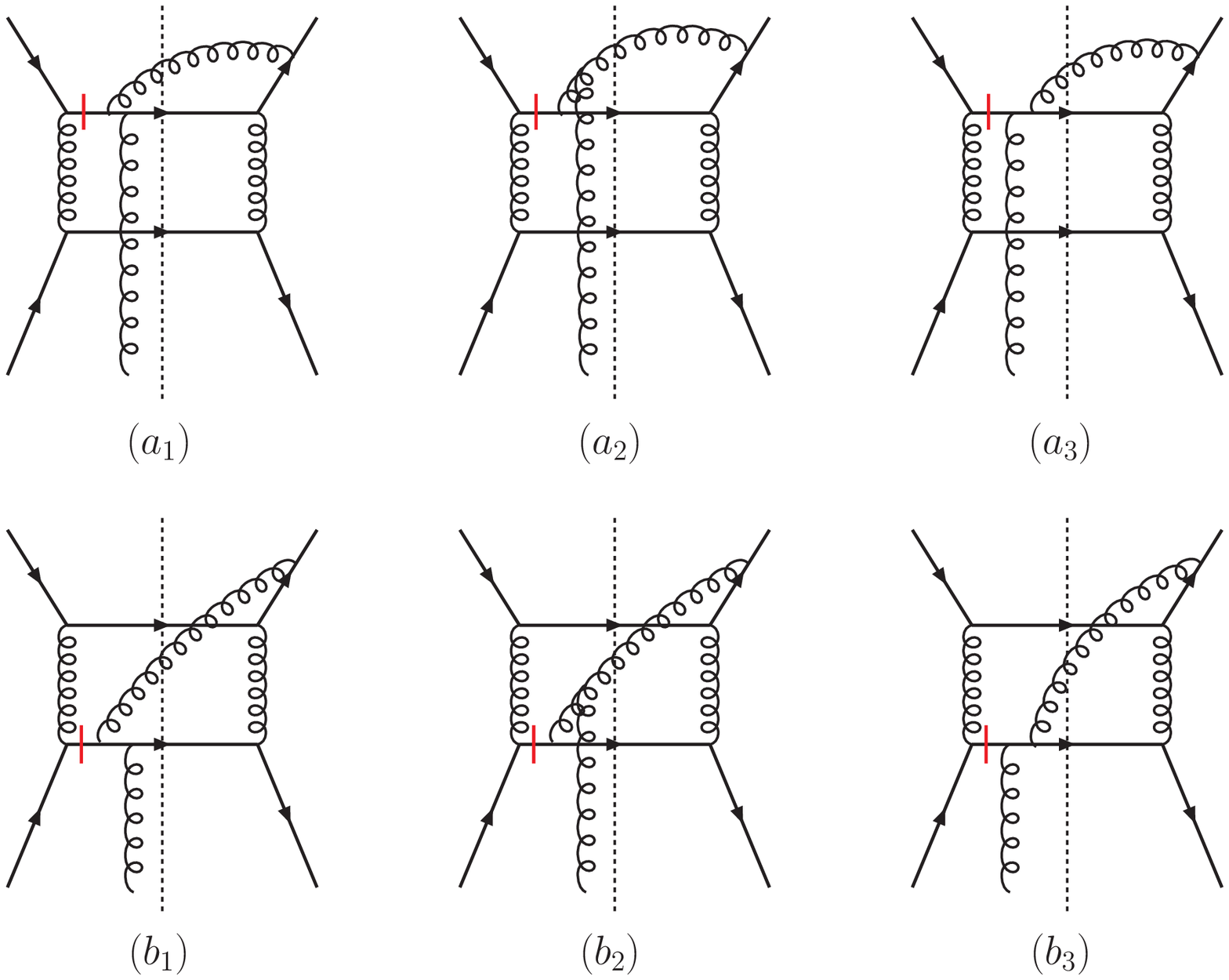}
\end{center}
\vskip -0.4cm \caption{\it Power suppressed hard pole
contributions when $k'$ is parallel to $P_B$.}\label{f22}
\end{figure}

The strong interaction phase necessary for a non-vanishing
$\Delta\sigma(S_\perp)$ arises from the interference between the
imaginary part of the partonic scattering amplitude with the extra
polarized gluon ($k_g=x_g P_A$) and the real scattering amplitude
without a gluon in Fig.~\ref{generic_diagrams}(b). The imaginary
part comes from taking the pole of parton propagator associated
with the integration over the gluon momentum fraction $x_g$.  For
a process with two physical scales, $P_\perp$ and $q_\perp$, tree
scattering diagrams in Fig.~\ref{generic_diagrams}(b) have two
types of poles, corresponding to $x_g=0$ (``soft-pole'')
\cite{qiusterman} and $x_g\neq 0$ (``hard-pole'')
\cite{JiQiuVogYua06}. When calculating the partonic scattering
amplitudes, we have to attach the polarized gluon to any
propagator of the hard part contained in the light circles in the
diagram of Fig.~\ref{generic_diagrams}(b). If the polarized gluon
attaches to the external quark lines either in the initial state
or in the final state, the on-shell propagation of the quark line
will generate a soft gluonic pole. A hard pole arises when
internal quark propagators go on-shell with nonzero $x_g$. In
Figs.~\ref{f15} and \ref{f20} we show the diagrams for the soft
pole contributions, and in Figs.~\ref{f18}-\ref{f22}
the ones for the hard pole contributions. In these figures, we only show
the diagrams with the additional gluon attaching to the left of the cut
line. Their mirror diagrams for which the gluon attaches to the right
are not shown, but are included in the final results. Certainly,
because of the additional gluon attachment, we will have many more
diagrams to calculate, compared to the calculations in the last
subsection for the unpolarized cross section. However, again, by
using the power counting technique, we are able to group all
diagrams according to, for example, whether the radiated gluon's
momentum $k'$ is nearly parallel to the polarized
hadron's momentum $P_A$, or to that of the unpolarized one, $P_B$.
Similar to the last subsection, we will focus on these
two types of contributions, because we want to investigate how
they may be factorized into the perturbatively generated TMD
parton distributions defined in Eqs.~(\ref{e36}),(\ref{e38}).
All other contributions are either associated with final-state effects, or
with the soft factor. In the evaluations of the
scattering amplitudes, we always keep only the leading power
contributions in $q_\perp/P_\perp$ and neglect all corrections that are
power-suppressed. In this way, we can clearly investigate how
the factorization works, and what the appropriate definition of
the TMD parton distributions is in this case.

The calculations of the soft-pole and hard-pole contributions to
the single spin asymmetry for the dijet-correlations follow the
same procedure as we used for the SIDIS and Drell-Yan processes
\cite{JiQiuVogYua06}, because they correspond to the same kinematic
limit, the intermediate transverse momentum region. In
Ref.~\cite{JiQiuVogYua06}, we first calculated the full
differential cross section valid for all transverse
momenta with $q_\perp\gg \Lambda_{\rm QCD}$, including the
region $q_\perp\sim Q$ (where $q_\perp$ is the transverse
momentum and $Q$ the virtuality of the photon in SIDIS or the
Drell-Yan process). The results were then
expanded in terms of $q_\perp/Q$ to obtain the leading power
contribution in the intermediate transverse momentum region. As
mentioned above, however, in this paper, we will utilize the power
counting technique from the beginning in order to simplify our
calculations. This means that our results are only valid in the
intermediate transverse momentum region $\Lambda_{\rm QCD}\ll q_\perp\ll
P_\perp$. A full calculation of the $q_T$ dependence of the dijet cross
section also at $q_\perp\sim Q$ would be extremely tedious and
not really provide any additional insights.
As a cross check, we have also recalculated the
Drell-Yan SSA in the intermediate transverse momentum region
following the method used in this paper, and found results identical to
our earlier ones in \cite{JiQiuVogYua06}.

The method for calculating the single transverse-spin asymmetry for
hard scattering processes in the twist-three approach has been
well developed and documented in the literature
\cite{qiusterman,JiQiuVogYua06,twist3-new,yuji06}. In the
following, we will just outline the main steps of the calculations
and highlight the unique features for this particular problem. For
further details of the twist-three calculations, we refer the reader to
\cite{qiusterman,JiQiuVogYua06,twist3-new,yuji06}. The collinear
expansion is the central step in obtaining the final results. We
perform our calculations in a covariant gauge. The additional gluon
from the polarized hadron is associated with a gauge
potential $A^\mu$, and one of the
leading contributions comes from its component $A^+$. Thus, the
gluon will carry longitudinal polarization. The gluon's
momentum is dominated by $x_gP_A+k_{g\perp}$, where $x_g$ is the
longitudinal momentum fraction with respect to the polarized
proton. The contribution to the single-transverse-spin asymmetry
arises from terms linear in $k_{g\perp}$ in the expansion of the
partonic scattering amplitudes. When combined with $A^+$, these
linear terms will yield $\partial^\perp A^+$, a part of the gauge
field strength tensor $F^{\perp +}$ in Eq.~(\ref{TF}). In this
collinear expansion, we further keep $k_{g\perp}\ll q_\perp$
because $q_\perp$ is a relative hard scale compared to
$k_{g\perp}$. Since $k_g=k_{2}-k_{1}$, the $k_{g\perp}$ expansion
of the scattering amplitudes can be performed through the
transverse momenta of $k_{1}$ and $k_{2}$, which we can
parameterize in the following way,
\begin{equation}
k_{1}=x_1P_A+k_{1\perp},~~~k_{2}=x_2P_A+k_{2\perp} \ , \label{e55}
\end{equation}
where we have neglected the minus components of the momenta,
as these do not contribute to the linear terms in the expansion in
$k_{g\perp}$. From momentum conservation, we know that
$k_{g\perp}=k_{2\perp}-k_{1\perp}$. Therefore, the collinear expansion in
$k_{g\perp}$ can be replaced by expansions in $k_{1\perp}$ and
$k_{2\perp}$:
\begin{equation}
{\cal H}_{(g)qq\to qqg}(k_{1\perp},k_{2\perp})={\cal H}_{(g)qq\to
qqg}(0,0)+k_{1\perp}^\rho\frac{\partial {\cal H}}{\partial
k_{1\perp}^\rho}|_{k_{1\perp}=k_{2\perp}=0}+k_{2\perp}^\rho\frac{\partial
{\cal H}}{\partial k_{2\perp}^\rho}|_{k_{1\perp}=k_{2\perp}=0} \ ,
\end{equation}
where ${\cal H}$ represents the amplitude squared for
the partonic process $(g)qq\to qqg$ including the delta function
for the phase space integral of $k'$, $\delta((k')^2)$ in~(\ref{x-ssa}).
Because of gauge invariance, the terms linear in $k_{1\perp}$ and
$k_{2\perp}$ can always be combined after summation over all diagrams,
\begin{equation}
\frac{\partial {\cal H}}{\partial
k_{1\perp}^\rho}|_{k_{1\perp}=k_{2\perp}=0} =-\frac{\partial {\cal
H}}{\partial k_{2\perp}^\rho}|_{k_{1\perp}=k_{2\perp}=0}\ .
\end{equation}
The final expression for the collinear expansion can then be written
as
\begin{equation}
{\cal H}_{(g)qq\to qqg}(k_{1\perp},k_{2\perp})={\cal H}_{(g)qq\to
qqg}(0,0)+ k_{g\perp}^\rho\frac{\partial {\cal H}}{\partial
k_{2\perp}^\rho}|_{k_{1\perp}=k_{2\perp}=0} \ .
\end{equation}
In order to obtain the complete result for the right-hand-side of the
above equation we have to keep track of the momentum flow in the
hard parts. One important contribution of the
$k_{g\perp}$ expansion comes from the on-shell condition for the
radiated gluon, whose momentum $k'$ depends on $k_{g\perp}$. This
leads to a term involving the {\it derivative} of the correlation
function $T_F$. In addition to the derivative contributions,
also non-derivative terms can arise from the
$k_{g\perp}$-expansion of other parts of the scattering
amplitudes. In the following, we will calculate the derivative and
non-derivative terms separately.

\subsubsection{Derivative Terms}

The derivative contribution is simpler. As summarized in
\cite{qiusterman}, the derivative terms come from two parts of the
collinear expansion. One is the on-shell condition for the
radiated gluon, $\delta((k')^2)$, mentioned above, and the other
is the double pole contribution in final state interactions. In
the derivation of these derivative contributions, we only need to
focus on these two parts in the $k_{g\perp}$ expansion of the
partonic amplitudes ${\cal H}$ and can neglect all other
contributions.

We first discuss the contributions by the on-shell condition $(k')^2=0$.
Since the $k_{g\perp}$ momentum flow will be different when the
polarized gluon attaches to the left or to the right of the cut, we
denote the radiated gluon's momentum as $k'_L$ or $k'_R$ in these
two cases. From momentum conservation, we find in
Fig.~\ref{f15}, for example, that $k_L'$ can be written as
\begin{equation}
k_L'=k_{2}+k_b-P_1-P_2=x_2P_A+k_{2\perp}+k_b-P_1-P_2\approx
k'+k_{2\perp}\ ,
\end{equation}
where $k'=xP_A+x'P_B-P_1-P_2$. Here, we have also approximated
$x_2$ by $x$; their difference will contribute to the
non-derivative terms but not to the derivative ones. With this
decomposition, $(k_L')^2$ becomes
\begin{equation}
(k_L')^2\approx
(k')^2-2k_{2\perp}\cdot(P_{1\perp}+P_{2\perp})=(k')^2-2k_{2\perp}\cdot
q_{\perp} \ ,
\end{equation}
where we have used the relation
$\vec{q}_\perp=\vec{P}_{1\perp}+\vec{P}_{2\perp}$. Here we have
also neglected higher-order terms in $k_{2\perp}$ which
do not contribute to the single-spin asymmetry. Similarly,
when the gluon attaches to the right, we find for
the on-shell condition:
\begin{equation}
(k_R')^2\approx (k')^2-2k_{1\perp}\cdot q_{\perp} \ .
\end{equation}
It is easy to see that the above two expansions differ only by
$k_{g\perp}\cdot q_\perp$. We know that the gluonic
poles have opposite signs for the two attachments, and that
apart from this difference the squared amplitudes for them are identical.
So we can combine the two contributions and find the
following result for the expansion of the delta function:
\begin{equation}
\delta((k_L')^2)-\delta((k_R')^2)=-2k_{g\perp}\cdot q_\perp
\frac{d\delta((k')^2)}{d(k')^2} \ .
\end{equation}
We can further rewrite the derivative of the delta function as a
derivative with respect to $x$,
\begin{equation}
\frac{d\delta((k')^2)}{d(k')^2}=\frac{d\delta((k')^2)}{dx}\frac{1}{d(k')^2/dx}
\ . \label{delta1}
\end{equation}
By using the momentum decomposition $k'=xP_A+x'P_B-P_1-P_2$, the
derivative $d(k')^2/dx_2$ can be written as
\begin{equation}
\frac{d(k')^2}{dx}=2k'\cdot P \ .\label{delta2}
\end{equation}
This leads to an important observation: the
derivative contributions will be power suppressed when the
radiated gluon's momentum $k'$ is parallel to the unpolarized
hadron's momentum $P_B$. This is because if $k'$ is parallel to
$P_B$, $2k'\cdot P_A$ will be of order $\hat s\sim P_\perp^2$,
and from Eq.~(\ref{delta1}) we can easily see that the
contribution will be suppressed by $1/P_\perp^2$. However, if $k'$
is parallel to $P_A$, $2k'\cdot P_A$ will be of order
$q_\perp^2$, instead of order $P_\perp^2$, so that in this case
the derivative contribution is not power suppressed. We can then
further parameterize $k'$ as
\begin{equation}
k^{\prime
\mu}=(1-\xi)xP_A^\mu+\frac{\vec{q}_\perp^2}{(1-\xi)2xP_A\cdot
P_B}P_B^\mu-q_\perp^\mu \ , \label{e66}
\end{equation}
where the coefficient in front of $P_B^\mu$ (the minus component
of $k'$) comes from the on-shell condition $(k')^2=0$, and where we
have also used the relation $k_\perp^{\prime \mu}=-q_\perp^\mu$.
Substituting the above into Eq.~(\ref{delta2}), we find that
\begin{equation}
\frac{d\delta((k')^2)}{dx}=\frac{\vec{q}_\perp^2}{x(1-\xi)} \ .
\end{equation}
This expression will appear in the final result for the
derivative terms.

When the polarized gluon attaches to the final state particles,
there will be double pole contributions to the single-spin
asymmetry \cite{qiusterman}. This is because the position of the
pole itself will depend on $k_{g\perp}$, and the expansion of the
associated propagator will lead to a double pole. For example, in the
final state interaction diagram of Fig.~\ref{f15}($b_1$), the propagator
giving a pole reads
\begin{equation}
\frac{1}{(P_1-k_g)^2+i\epsilon}=\frac{1}{(P_1-x_gP_A)^2-2P_{1\perp}\cdot
k_{g\perp}+i\epsilon} \ . \label{poleprop}
\end{equation}
The expansion in $k_{g\perp}$ will give a double pole. This double
pole will make a contribution to the derivative terms \cite{qiusterman}.
However, it turns out that
this double pole contribution to the derivative terms is
power suppressed in the intermediate transverse momentum region
$\Lambda_{\rm QCD}\ll q_\perp\ll P_\perp$. In order to
demonstrate this, we will follow the method described
in \cite{twist3-new} to calculate the double pole contributions to
the derivative terms. We recall that it is the imaginary part of the
above propagator that contributes to the single-spin asymmetry. We can
first evaluate the imaginary part and perform the $k_{g\perp}$
expansion afterwards. The imaginary part is proportional to the
pole of the propagator in~(\ref{poleprop}),
$\delta((P_1-k_g)^2)$, for which the $k_{g\perp}$ expansion gives
\begin{equation}
\delta((P_1-k_g)^2)=\delta((P_1-x_gP_A)^2)+\delta'((P_1-x_gP_A)^2)
(-2P_{1\perp}\cdot
k_{g\perp}) \ . \label{e69}
\end{equation}
For the derivative contribution, the double pole will have the
same form as that for the expansion of the on-shell condition for
the radiated gluon discussed above. When we convert the derivative
of the delta function in Eq.~(\ref{e69}) to a derivative with respect
to $x_g$, we find a suppression factor proportional to
\begin{equation}
\frac{1}{2P_1^-P_A^+}\sim \frac{1}{P_\perp^2} \ .
\end{equation}
Therefore, this double pole contribution is power suppressed in
the $q_\perp/P_\perp$ expansion. Similarly, the double pole
contribution to the derivative terms from the final state
interaction on the line with momentum $P_2$ like the one shown in
Fig.~\ref{f15}($c_1$) is also power suppressed.

\begin{figure}[t]
\begin{center}
\includegraphics[width=12cm]{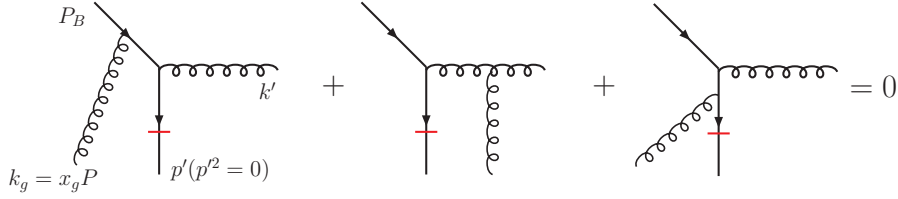}
\end{center}
\vskip -0.4cm \caption{\it There is no contribution to the
derivative terms from the hard poles because of the Ward
identity.}\label{f16}
\end{figure}

Another important observation is that the derivative term in the
hard pole contributions vanishes when all diagrams are summed. This
has been shown in the explicit calculations for the Drell-Yan and
SIDIS processes \cite{JiQiuVogYua06}, and has also been
demonstrated from a more general argument based on the analytic
property of the momentum expansion and on gauge invariance of the
twist-three matrix elements \cite{yuji06}. Here, we will give
another proof for this result based on the Ward identity. Take one
of the hard pole contributions shown in Fig.~\ref{f18}(a) as an
example, where all three diagrams ($a_1$-$a_3$) contribute to the
same hard pole as indicated by the bars in the propagators. As
discussed above, the derivative terms come from either the
expansion of the delta function for the on-shell condition for the
radiated gluon, or from the double pole in the final state interaction
(which also corresponds to a delta function as we showed above). When we
expand the delta functions, all other parts of the amplitudes will be
evaluated at $k_{g\perp}=0$. Comparing these three diagrams
$a_1$-$a_3$, we find that the only difference exists in the vertices
where the polarized gluon attaches to the incoming quark, outgoing
gluon, or the quark propagator, while the rest of the squared
amplitudes is the same in each case. Thus, we can separate off this
vertex part, and sum the remaining parts of the diagrams. These three
vertex parts form a gauge-invariant set of lowest-order $qg\to qg$
scattering diagrams with all external partons on mass shell.
Without $k_{g\perp}$-flow in these vertices, the attaching gluon
only has longitudinal momentum ($k_g=x_gP_A$), and its
polarization is also along the longitudinal direction.  From the
Ward identity, the sum of these three diagrams vanishes because
all other partons in this lowest-order scattering amplitude are on
mass shell and gluons have physical polarization. The situation is
described in Fig.~\ref{f16}.

\begin{table}[t]
\caption{The color factors for the diagrams in Fig.~\ref{f15}. }
\begin{ruledtabular}
\begin{tabular}{|l|c|c|c|c|c|}
& (1) & (2) & (3) & (4) & (5)   \\
\hline $a$  & $\frac{1}{2N_c}\frac{1}{2N_c^2}$
&$\frac{1}{2N_c}\frac{1}{2N_c^2}$
       &0 & $-\frac{1}{2N_c}\frac{1}{2N_c}$& $-\frac{1}{2N_c}\frac{1}{2N_c}$\\
\hline $b$ & $\frac{1}{2N_c}\frac{1}{4N_c^2}$ &
     $\frac{1}{2N_c}\frac{N_c^2+1}{4N_c^2}$
       & $-\frac{1}{2N_c}\frac{1}{4}$ &$-\frac{1}{2N_c}
\frac{N_c^2-2}{4N_c^2}$ & $\frac{1}{2N_c}\frac{1}{2N_c^2}$\\
\hline $c$ & $-\frac{1}{2N_c}\frac{N_c^2-2}{4N_c^2}$ &
$-\frac{1}{2N_c}\frac{N_c^2-2}{4N_c^2}$ & 0 &
$-\frac{1}{2N_c}\frac{1}{2N_c}$& $-\frac{1}{2N_c}\frac{1}{2N_c}$
 \end{tabular}
\end{ruledtabular}
\end{table}

\begin{figure}[t]
\begin{center}
\includegraphics[width=12cm]{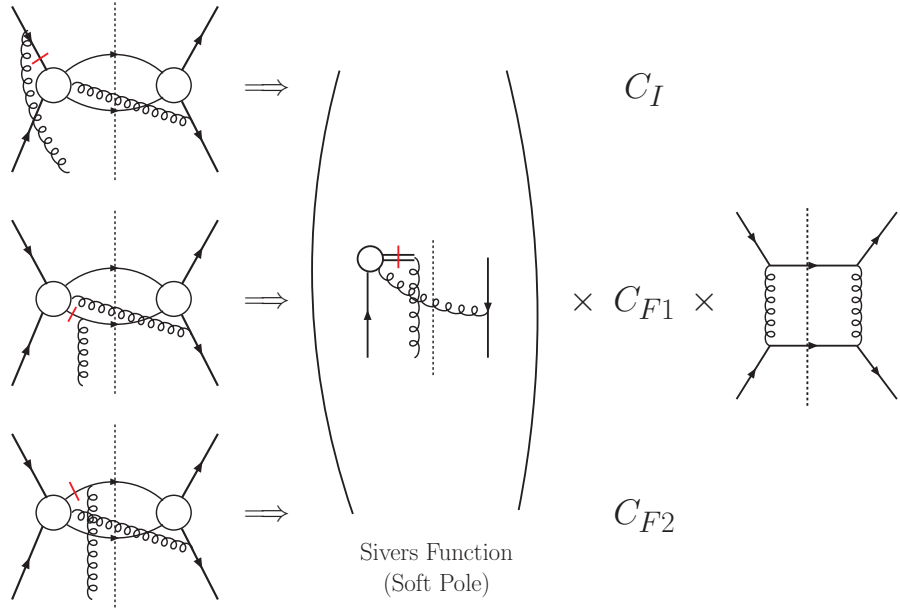}
\end{center}
\vskip -0.4cm \caption{\it Illustration of the factorization of
the soft pole contributions when $k'$ is parallel to $P_A$. This
part can be factorized into a soft pole contribution to the
perturbatively generated Sivers function in polarized hadron
$P_A$. This factorization applies both to the derivative and
the non-derivative contributions from the soft poles. The circle with
the attached gluon in the Feynman diagrams represents all possible
gluon attachments. Likewise, the gluon attachment to
the circle in the Sivers function diagram includes the diagrams
with gluon attachments to the gauge link and to the quark line.}
\label{f17}
\end{figure}

In summary, in the intermediate transverse momentum region
$\Lambda_{\rm QCD}\ll q_\perp\ll P_\perp$, we will only have
leading-power contributions to the derivative terms from the
expansion of the delta function for the on-shell condition of the
radiated gluon $k'$, and when $k'$ is parallel to $P_A$. We have
shown these diagrams in Fig.~\ref{f15}. This part of the
contribution is relatively easy to derive, as we mentioned above,
by multiplying a factor $(1-\xi)x/\vec{q}_\perp^2$ to the
unpolarized cross sections. The only additional involvement is the
color factor. We list the color-factors for all diagrams of
Fig.~\ref{f15} in Table IV. Each of the color factors has a factor
$1/2N_c$. This can be easily seen for the color factors for
diagrams $a_1$, $b_1$ and $c_1$, for which we have the following
simplifications,
\begin{eqnarray}
a_1&:& \frac{1}{N_c}\frac{1}{C_F}{\rm Tr}(T^bT^aT^bT^cT^d){\rm
Tr}(T^aT^cT^d)
{\rm Tr}(T^aT^cT^d)
=-\frac{1}{2N_c}\times
C_I\nonumber\ , \\
b_1&:& \frac{1}{N_c}\frac{1}{C_F}{\rm Tr}(T^bT^aT^bT^cT^aT^d){\rm
Tr}(T^cT^d)
{\rm Tr}(T^cT^d)
=-\frac{1}{2N_c}\times C_{F_1}\nonumber\ , \\
c_1&:& \frac{1}{N_c}\frac{1}{C_F}{\rm Tr}(T^bT^aT^bT^cT^d){\rm
Tr}(T^aT^dT^c)
{\rm Tr}(T^aT^cT^d)
=-\frac{1}{2N_c}\times C_{F2}\ ,
\end{eqnarray}
where $C_I$, $C_{F1}$ and $C_{F2}$ have been defined above in
Eqs.~(\ref{e25}),(\ref{e26}),(\ref{e27}). Combining these color factors
with the squared amplitude calculated in the last
subsection for the unpolarized cross section, we find that the
contributions to the derivative terms can be factorized into the
hard factor given in Eq.~(\ref{hqq}) times the derivative terms of the
perturbatively generated Sivers function for the polarized
nucleon in Eq.~(\ref{e38}). We
illustrate this factorization in Fig.~\ref{f17}. For example, the
derivative contribution from diagram ($a_1$) is factorized into a
soft pole diagram contribution to the derivative term in the
Sivers function where the radiated gluon attaches to the quark
line shown in Fig.~\ref{f17}, multiplied by the color-factor $C_I$
and the partonic amplitude squared for $qq'\to qq'$. $b_1$ and
$c_1$ will be factorized into the same diagram contribution to the
quark Sivers function, but with different color-factors $C_{F1}$
and $C_{F2}$, respectively. The same conclusion holds for all
other diagrams. For example, the sum of diagrams ($a_2$-$a_5$) is
factorized into a soft pole diagram contribution to the Sivers
function where the radiated gluon attaches to the gauge link,
multiplied by the color factor $C_I$ and the partonic amplitude
squared for $qq'\to qq'$.

Finally, we present the total result for the derivative
contribution by the $qq'$ process to the single-spin asymmetry for the
dijet-correlation,
\begin{eqnarray}
\left.
\frac{d\Delta\sigma(S_\perp)_{(qq')}^{\rm D}}
     {dy_1dy_2dP_\perp^2d^2\vec{q}_\perp}
\right|_{k'\propto P_A}
&=&
H_{qq'\to qq'}^{\rm Sivers}(\hat{s},\hat{t},\hat{u})\,
\frac{\epsilon^{\alpha\beta}S_\perp^\alpha q_\perp^\beta}
     {(\vec{q}_\perp^2)^2}\,
 \left[x_b\,q'(x_b)\right]
\nonumber\\
&\times&
\frac{\alpha_s}{2\pi^2}\,
\left[-\frac{1}{2N_c}\right]
\int dx\, \xi
\left[x\frac{\partial}{\partial x}T_F(x,x)\right]
\left[1+\xi^2\right] \, ,
\label{x-ssa-d}
\end{eqnarray}
where the superscript ``D'' indicates the derivative term, and where
the single-scale partonic hard part $H_{qq'\to qq'}^{\rm Sivers}$ has been
defined in Eq.~(\ref{hqq}). The result in Eq.~(\ref{x-ssa-d})
indeed reproduces the derivative terms in the first order TMD
factorization formula in Eq.~(\ref{e39}).

\subsubsection{Non-derivative terms: $k'$ parallel to $P_A$}

In addition to the derivative contributions, the soft poles also
contribute to the non-derivative terms. One way of calculating the
non-derivative terms is to follow the $k_{g\perp}$-flow in
the scattering amplitude squared, summing up all contributions.
Keeping in mind that $\vec{k}_{g\perp}=\vec{k}_{2\perp}-\vec{k}_{1\perp}$,
we have to keep the full dependence on $k_{2\perp}$ and $k_{1\perp}$
in the calculations. It is easy to keep track of the explicit $k_{i\perp}$
dependence coming from momentum conservation. In addition, we also
need to include the ``indirect'' $k_{i\perp}$ dependence resulting
from kinematic constrains: the on-shell condition of the radiated gluon
and the double poles in the final state interactions. For example,
the on-shell delta function for the momentum $k'$ of the radiated
gluon requires that the longitudinal momentum component depends on
$k_{g\perp}$, which will lead to a $k_{g\perp}$ dependence for the
longitudinal momentum fractions of $k_{1}$ and $k_{2}$ as well.

As mentioned above, if the polarized gluon attaches to the left
side of the cut, the radiated gluon will have momentum
$k_L'=k_{2}+k_b-P_1-P_2$. Since $k_{2}$ has transverse momentum
$k_{2\perp}$, the transverse component of $k_L'$ will be equal to
$\vec{k}_{L\perp}'=\vec{k}_{2\perp}-\vec{P}_{1\perp}-
\vec{P}_{2\perp}=\vec{k}_{2\perp}-\vec{q}_\perp$. Furthermore, the
minus component of $k_L'$ is fixed by momentum conservation to
$k_L^{\prime -}\approx k_b^--P_1^--P_2^-$, where the contribution
by $k_{2}^-$ has been neglected, because it is of order ${\cal
O}(k_{2\perp}^2)$. Therefore, $k_L^{\prime -}$ does not depend
linearly on $k_{g\perp}$. The plus component of $k_L'$, on the
other hand, will have linear $k_{g\perp}$ dependence because of
the on-shell condition. We therefore parameterize $k_L'$ as
\begin{equation}
k_L^{\prime\mu}=(1-\xi)x\left(1+{\cal
O}(k_{2\perp})\right)P_A^\mu+\beta
P_B^\mu+k_{2\perp}^\mu-q_{\perp}^\mu \ ,
\end{equation}
where we have indicated the term linear in $k_{2\perp}$ in
the plus component of $k_L'$. As mentioned above, the
minus component $\beta$ does not depend linearly on
$k_{g\perp}$ and can be determined to be
$\beta=\vec{q}_\perp^2/2(1-\xi)xP_A\cdot  P_B$. By using the
on-shell condition $(k_L')^2=0$, we can then obtain the
$k_{g\perp}$-dependent term in the plus component of $k_L'$, and
the final parameterization result for $k_L'$ is
\begin{equation}
k_L^{\prime \mu}=(1-\xi)x\left(1-\frac{2q_\perp\cdot
k_{2\perp}}{q_\perp\cdot
q_\perp}\right)P_A^\mu-\frac{-\vec{q}_\perp^2}{2(1-\xi)xP_A\cdot
P_B}P_B^\mu+k_{2\perp}^\mu-q_\perp^\mu \ . \label{e75}
\end{equation}
It is easy to check that $(k_L')^2=0$ up to terms quadratic in
$k_{2\perp}$. From this result, we can also determine the
$k_{2\perp}$-dependence of the plus components of $k_{1}$ and
$k_{2}$. For example, $k_{2}$ can be
directly calculated from momentum conservation, and we find
\begin{equation}
k_{2}^\mu=x\left[1-(1-\xi)\frac{2q_\perp\cdot
k_{2\perp}}{q_\perp\cdot q_\perp}\right]P_A^\mu+k_{2\perp}^\mu \ ,
\label{e76}
\end{equation}
where the minus component of $k_{2}$ has again been
neglected. We next obtain the
parameterization for $k_{1}$ by $k_{1}=k_{2}-k_{g}$, where $k_g$
is determined by the pole and will be different for different
poles. For the initial state interaction, the pole
$(P_B+k_g)^2=0$ gives $k_g^\mu=k_{g\perp}^\mu$, and the
longitudinal momentum fraction does not depend on the linear term
in $k_{g\perp}$. However, for the final state interaction with
gluon attachment to the line with momentum $P_1$, the pole
will be given by $(P_1+k_g)^2=0$, and
the plus component of $k_g$ will depend on $k_{g\perp}$ through
$k_g^\mu=-\frac{P_1\cdot k_{g\perp}}{P_A\cdot
P_1}P_A^\mu+k_{g\perp}^\mu$. Similarly, the final state
interaction with $P_2$ will lead to $k_g^\mu=-\frac{P_2\cdot
k_{g\perp}}{P_A\cdot  P_2}P_A^\mu+k_{g\perp}^\mu$. In this way, the
parameterizations for $k_{1}$ and $k_g$ will vary for
different poles, whereas that for $k_{2}$ will always remain the same.
This is because the parameterization of $k_{2}$ comes from the
on-shell condition for the radiated gluon, which is not sensitive to
the location of the poles.

When the polarized gluon attaches to the right of the cut
lines, on the other hand, it is $k_{1}$ that is determined by the
on-shell condition for the radiated gluon, $(k_R')^2=0$. Following
the same method as above, we find
\begin{eqnarray}
k_R^{\prime \mu}&=&(1-\xi)x\left(1-\frac{2q_\perp\cdot
k_{1\perp}}{q_\perp\cdot
q_\perp}\right)P_A^\mu-\frac{-\vec{q}_\perp^2}{2(1-\xi)xP_A\cdot P_B}P_B^\mu+k_{1\perp}^\mu-q_\perp^\mu \nonumber\ ,\\
k_{1}^\mu &=& x\left[1-(1-\xi)\frac{2q_\perp\cdot
k_{1\perp}}{q_\perp\cdot q_\perp}\right]P_A^\mu+k_{1\perp}^\mu \ .
\end{eqnarray}
Furthermore, in this case $k_{2}=k_{1}+k_g$, with $k_g$ determined
from the position of the poles; $k_{2}$ will be different for the
different poles.

With the above parameterizations for all the external particles'
momenta, the calculations for the linear expansion in $k_{g\perp}$
are straightforward. The final result for the soft pole
contributions to the non-derivative terms when $k'$ is nearly
parallel to $P_A$ have the following form:
\begin{eqnarray}
\left.
\frac{d\Delta\sigma(S_\perp)_{(qq')}^{\rm soft}}
     {dy_1dy_2dP_\perp^2d^2\vec{q}_\perp}
\right|_{k'\propto P_A}
&=&
H_{qq'\to qq'}^{\rm Sivers}\,
\frac{\epsilon^{\alpha\beta}S_\perp^\alpha q_\perp^\beta}
     {(\vec{q}_\perp^2)^2}\,
 \left[x_b\,q'(x_b)\right]
\nonumber\\
&\times&
\frac{\alpha_s}{2\pi^2}\,
\left[-\frac{1}{2N_c}\right]
\int dx\, \xi \,
T_F(x,x)
\left[\frac{2\xi^3-3\xi^2-1}{1-\xi}\right]\, ,
\label{x-ssa-nd-sa}
\end{eqnarray}
where $H_{qq'\to qq'}^{\rm Sivers}$ is given in Eq.~(\ref{hqq}).
The separate factorization for initial and final state
interactions diagrams in Fig.~\ref{f17}, which we used to describe
the factorization of the derivative contributions, also applies to
the non-derivative ones. For example, the contributions from
diagrams ($a_1$-$a_5$) of Fig.~\ref{f15} can be factorized into
the Sivers function shown in Fig.~\ref{f17}, multiplied by $C_I$
and the partonic scattering function, including both the
derivative and non-derivative contributions. The same happens for the
final state interaction diagrams ($b_1$-$b_5$) and ($c_1$-$c_5$).

There are also hard pole contributions when $k'$ is parallel to
$P_A$. We have shown these diagrams in Fig.~\ref{f18}. As discussed
above, the hard pole diagrams only contribute to the
non-derivative terms. To calculate their contributions, we follow
the above method as for the soft pole contributions. Because the hard
poles appear in the internal propagator, we have to take {\it all}
possible gluon attachments into account. For example, diagrams
$a_1$-$a_3$ of Fig.~\ref{f18} display all contributions with the
same hard pole (indicated by the short bars in the propagators),
with three different attachments of the polarized gluon.

Another important feature for the hard pole contributions is that,
after taking the pole, the polarized gluon will have nonzero
longitudinal momentum fraction, i.e., $x_g\neq 0$ as we mentioned
above. This is in fact the reason for referring to these poles as
``hard'' poles. For example, the hard poles in the diagrams
$a_1$-$a_3$ have the momentum $k_b-k'+k_g$ with $k_g=x_gP_A$, and
their pole arises at the following value for $x_g$:
\begin{equation}
x_g=\frac{-(k_b-k')^2}{2P_A\cdot  (k_b-k')}\approx \frac{2k_b\cdot
k'}{2k_b\cdot P_A}=(1-\xi)x \ ,
\end{equation}
where the last equality comes from the fact that $k'$ is
parallel to $P_A$ and is dominated by its plus component as shown
in Eq.~(\ref{e66}). We have also neglected all higher power
corrections in $q_\perp^2/P_\perp^2$ in the above derivations.
From this result, we find the associated twist-three quark-gluon
correlation function will take the form $T_F(x,x-x_g)=T_F(x,\xi
x)$, that is, its two arguments are not at the same. This is
different from the soft pole contributions where $x_1$ and $x_2$
are equal in the twist-3 quark-gluon correlation matrix element.

\begin{figure}[t]
\begin{center}
\includegraphics[width=12cm]{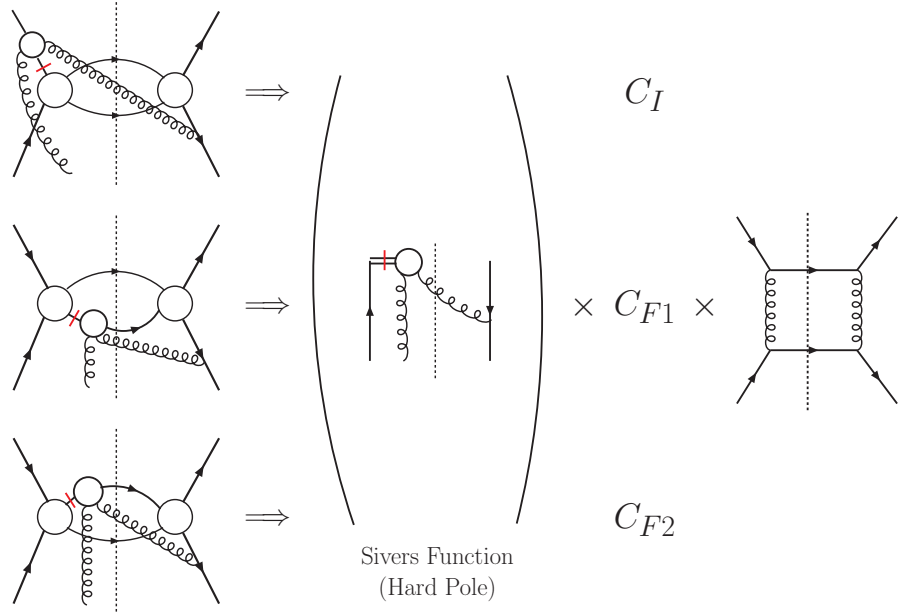}
\end{center}
\vskip -0.4cm \caption{\it Illustration of the factorization of
the hard pole contributions when $k'$ is parallel to $P_A$. This
part can be factorized into a hard pole contribution to the Sivers
function in the polarized hadron $A$. The little circle in the
diagrams represents all possible gluon attachments, as required by
gauge-invariance. For example, the first one in the left
panel represents all three diagrams $a_1$-$a_3$ of
Fig.~\ref{f18}.}\label{f19}
\end{figure}

In our calculations, we further choose physical polarization for
the radiated gluon, with the following polarization tensor:
\begin{equation}
\sum_\lambda \epsilon_\lambda^\mu(k')
\epsilon_\lambda(k')^{\mu'*}=-g^{\mu\mu'}+
\frac{k^{\prime\mu}P_B^{\mu'}+k^{\prime\mu'}P_B^{\mu}}{k'\cdot
P_B} \ .
\end{equation}
With this choice, we only have contributions from the
three-gluon vertex diagrams $a_2$, $b_2$ and $c_2$
and, apart from the color factor, each of them
contributes a term $\frac{1+\xi}{1-\xi}$. Their
color factors are straightforward to decompose:
\begin{eqnarray}
a_2&:& \frac{1}{N_c}\frac{1}{C_F}(-if_{ade}){\rm
Tr}(T^aT^dT^bT^c){\rm
Tr}(T^eT^bT^c)
=\left(C_F+\frac{1}{2N_c}\right)\times C_I \nonumber\
,\\
b_2&:& \frac{1}{N_c}\frac{1}{C_F}(-if_{ade}){\rm
Tr}(T^aT^dT^bT^eT^c){\rm
Tr}(T^bT^c)
=\left(C_F+\frac{1}{2N_c}\right)\times C_{F1} \nonumber\
,\\
a_2&:& \frac{1}{N_c}\frac{1}{C_F}(-if_{ade}){\rm
Tr}(T^aT^dT^bT^c){\rm
Tr}(T^eT^cT^b)
=\left(C_F+\frac{1}{2N_c}\right)\times C_{F2} \ .
\end{eqnarray}
We then directly find that the contributions of these diagrams
can be factorized into the hard pole contributions
to the Sivers function, multiplied by $C_I$ (or $C_{F1}$ and
$C_{F2}$) and the partonic hard-scattering function. We display this
factorization graphically in Fig.~\ref{f19}. The total contribution is
given by
\begin{eqnarray}
\left.
\frac{d\Delta\sigma(S_\perp)_{(qq')}^{\rm hard}}
     {dy_1dy_2dP_\perp^2d^2\vec{q}_\perp}
\right|_{k'\propto P_A} &=& H_{qq'\to qq'}^{\rm Sivers}\,
\frac{\epsilon^{\alpha\beta}S_\perp^\alpha q_\perp^\beta}
     {(\vec{q}_\perp^2)^2}
 \left[x_b\,q'(x_b)\right]
\nonumber\\
&\times&
\frac{\alpha_s}{2\pi^2}\,
\left[-\left(\frac{1}{2N_c} + C_F\right)\right]
\int dx\, \xi \,
T_F(x,x-\hat x_g)
\left[\frac{1+\xi}{1-\xi}\right] \ ,
\label{x-ssa-nd-ha}
\end{eqnarray}
where $\hat{x}_g=(1-\xi)x$ and $H_{qq'\to qq'}^{\rm Sivers}$ is
given in Eq.~(\ref{hqq}). Indeed, the above result reproduces the
contribution to Eq.~(\ref{e39}) that arises from the hard pole
part of the quark Sivers function.

In summary, by combining all the soft-pole and hard-pole
contributions when the radiated gluon is nearly parallel to the polarized
nucleon, we can reproduce the contribution from the perturbatively
generated Sivers function to the SSA in dijet-correlations in
Eq.~(\ref{e39}), derived from the TMD factorization formula in the
regime $\Lambda_{\rm QCD}\ll q_\perp\ll P_\perp$. With these
explicit calculations, we have demonstrated that at this perturbative
order the contribution to the SSA when $k'$ is parallel to $P_A$
can be factorized into the quark Sivers function, multiplied by
the initial/final state interaction color-factors ($C_I$, $C_{F1}$,
and $C_{F2}$), and the partonic scattering amplitude squared.

%

\subsubsection{Non-derivative terms: $k'$ parallel to $P_B$}

The calculation for the contributions when $k'$ is parallel to
$P_B$ follows the same procedure as above. As we have shown earlier,
there is no contribution to the derivative terms when
$k'$ is parallel to $P_B$, and so we only need to consider the
non-derivative contributions. We will also have both soft and hard
poles contributions. In Fig.~\ref{f20}, we have shown the soft pole
diagrams, and in Figs.~\ref{f21},\ref{f22} the hard pole ones.

Again we have to keep track of the $k_{g\perp}$-flow in the scattering
amplitudes. First, we need to parameterize the momenta of all the external
particles including their linear dependence on $k_{g\perp}$. For
example, when the polarized gluon attaches to the left of the cut line
in the diagrams of Figs.~\ref{f20}-\ref{f22}, the momentum $k_L'$ of the
radiated gluon and the quark momentum $k_{2}$ can be parameterized as
\begin{eqnarray}
k_L^{\prime \mu}&=&(1-\xi')x'P_B^\mu+\frac{2k_{2\perp}\cdot
q_\perp-q_\perp\cdot q_\perp}{2x'(1-\xi')P_B\cdot
P_A}P_A^\mu+k_{2\perp}^\mu-q_\perp^\mu \ ,\nonumber\\
k_{2}^\mu &=& x\left(1+\frac{2k_{2\perp}\cdot
q_\perp}{2(1-\xi')xx'P_A\cdot  P_B}\right)P_A^\mu+k_{2\perp}^\mu \
,
\end{eqnarray}
and $k_{1}=k_{2}-k_g$ with $k_g$ determined by the pole. Note that
these parameterizations are different from those in
Eqs.~(\ref{e75}),(\ref{e76}). This is because, when $k'$ is parallel
to $P_B$, its minus component is dominant and is parameterized by
$(1-\xi')x'P_B^\mu$. Again, this minus component does not depend
linearly on $k_{g\perp}$. The plus component of $k_L'$ does have
linear dependence, which can be calculated from the on-shell condition
$(k_L')^2=0$. The linear $k_{g\perp}$-dependence for the plus
component of $k_{2}$ can be calculated as well.

Likewise, when the gluon attaches to the right of the cut, we will have
\begin{eqnarray}
k_R^{\prime \mu}&=&(1-\xi')x'P_B^\mu+\frac{2k_{1\perp}\cdot
q_\perp-q_\perp\cdot q_\perp}{2x'(1-\xi')P_B\cdot
P_A}P_A^\mu+k_{1\perp}^\mu-q_\perp^\mu \ ,\nonumber\\
k_{1}^\mu &=&x\left(1+\frac{2k_{1\perp}\cdot
q_\perp}{2(1-\xi')xx'P_A\cdot  P_B}\right)P_A^\mu+k_{1\perp}^\mu \
,
\end{eqnarray}
and $k_{2}=k_{1}+k_g$ with $k_g$ determined from the position of the
poles. Again, different poles will give different results for $k_g$,
following the above discussions for the soft pole contributions.

\begin{figure}[t]
\begin{center}
\includegraphics[width=12cm]{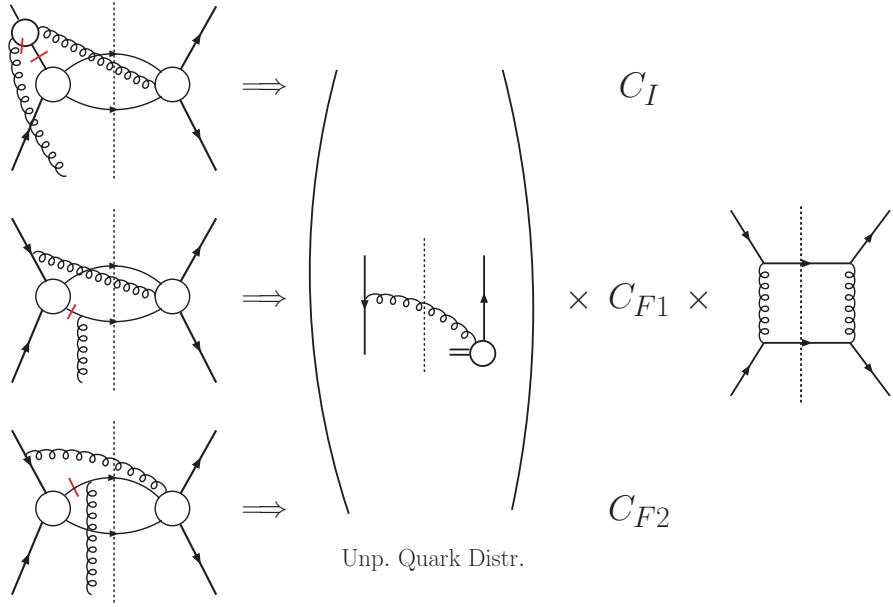}
\end{center}
\vskip -0.4cm \caption{\it Factorization of the contributions when
$k'$ is parallel to $P_B$. This contribution is factorized
into the TMD quark distribution in the unpolarized nucleon $B$.
The little circle in the diagrams represents all possible gluon
attachments. The bars indicate the poles.
The second and third diagrams in the left panel
represent the soft pole diagrams $b_1$-$b_5$ and $c_1$-$c_5$ of
Fig.~\ref{f20} respectively. The first one contains both soft poles
(represented by the bar on the circle (diagrams
$a_1$-$a_5$ of Fig.~\ref{f20})) and hard poles (indicated by the bar on the
straight line (all diagrams in Fig.~\ref{f21})).}\label{f23}
\end{figure}

By using the above parameterizations for the momenta of the
external particles, we can calculate the contributions from the
soft pole diagrams in Fig.~\ref{f20}, and the hard pole diagrams
in Figs.~\ref{f21} and \ref{f22}. We further find that the
contribution from the hard pole diagrams in Fig.~\ref{f22} is
power suppressed by $q_\perp/P_\perp$. So, for the hard pole
contributions, we only need to consider the diagrams shown in
Fig.~\ref{f21}, which all have the same hard pole indicated by the
bars on the propagators. As discussed in the last subsection, the
hard pole contributions may in general lead to the
polarized gluon having nonzero longitudinal momentum fraction,
$x_g\neq 0$. However, when the radiated gluon's momentum $k'$ is
nearly parallel to $P_B$, the hard poles lead to $x_g=0$
at leading power, as was the case for the soft pole contributions.
This is very different from what we found for
the hard pole contributions in the last subsection in the case
$k'$ parallel to $P_A$. The momentum giving the hard pole, $x'P_B-k'+k_g$
with $k_g=x_gP_A$, is the same as in the last subsection. However,
taking the pole, we now find
\begin{equation}
x_g=\frac{-(x'P_B-k')^2}{2k_b\cdot P_A}\approx 0 \ , \label{xdef}
\end{equation}
where the last equation holds in the limit $q_\perp/P_\perp\to
0$. The reason for this is that, when $k'$ becomes parallel to
$P_B$, it will be dominated by its minus component and can be
parameterized as
$k^{\prime\mu}=(1-\xi')x'P_B^\mu+\vec{q}_\perp^2
P_A^\mu/2(1-\xi')x'P_B\cdot P_A+k_\perp^{\prime\mu}$. Substituting
this into the above equation~(\ref{xdef}), we find that
the numerator is of order $q_\perp^2$, whereas the denominator is of
order $P_\perp^2$.
So, in the leading power approximation, $x_g\approx 0$ for the
hard pole contributions when $k'$ is nearly parallel to $P_B$. The
contributions from the soft poles in Fig.~\ref{f20} and the hard
poles in Fig.~\ref{f21} is added to obtain the final
results. We emphasize again that our power-counting based
analysis is only valid for the leading power terms in the
$q_\perp/P_\perp$ expansion.

The final result for the contributions from Figs.~\ref{f20} and
\ref{f21}, plus their mirror diagrams, is
\begin{eqnarray}
\left.
\frac{d\Delta\sigma(S_\perp)_{(qq')}}
     {dy_1dy_2dP_\perp^2d^2\vec{q}_\perp}
\right|_{k'\propto P_B} &=& -H_{qq'\to qq'}^{\rm
Sivers}(\hat{s},\hat{t},\hat{u})\,
\frac{\epsilon^{\alpha\beta}S_\perp^\alpha q_\perp^\beta}
     {(\vec{q}_\perp^2)^2}
 \left[x_a\,T_F(x_a,x_a)\right]
\nonumber\\
&\times&
\frac{\alpha_s}{2\pi^2}\,
C_F \,
\int dx'\, \xi' \,
q'(x') \left[\frac{1+\xi^{\prime 2}}{1-\xi'}\right] \ ,
\label{x-ssa-nd-b}
\end{eqnarray}
which is again proportional to the same hard factor $H_{qq'\to
qq'}^{\rm Sivers}$, and reproduces the last term in the TMD
factorization formula Eq.~(\ref{e39}). It is important to note
that a part of the hard pole contributions from Fig.~\ref{f21}
cancels out the soft pole contributions from diagrams $a_1$-$a_5$
of Fig.~\ref{f20}. The remainder of the hard pole contribution is
factorized into the quark distribution multiplied by $C_I$ and the
partonic scattering function. The soft pole diagrams $b_1$-$b_5$
and $c_1$-$c_5$ are separately factorized into the same quark
distribution for the unpolarized nucleon multiplied by $C_{F1}$
and $C_{F2}$, respectively, and the partonic scattering function.
We illustrate this factorization in Fig.~\ref{f23}.

After summing up all contributions from Eqs.~(\ref{x-ssa-d}),
(\ref{x-ssa-nd-sa}), (\ref{x-ssa-nd-ha}), and (\ref{x-ssa-nd-b}),
we obtain the total leading-power contribution to
$\Delta\sigma(S_\perp)$ from the $(g)qq'\to qq'g$ partonic
subprocess in the $q_\perp/P_\perp$ expansion. It reproduces the
factorized formula in Eq.~(\ref{e39}), which is the leading term
in the corresponding TMD factorization formula, Eq.~(\ref{e4}).

In summary, we have calculated the single transverse-spin asymmetry in
dijet-correlation in the twist-three approach when the radiated
gluon is parallel to either the polarized nucleon or the
unpolarized nucleon. By using the power counting technique, we
have shown that the contribution to the SSA can be factorized into
the perturbatively generated quark Sivers function when the
radiated gluon is parallel to the polarized nucleon, and into the
unpolarized quark distribution when it is parallel to the
unpolarized nucleon. We have demonstrated that the
result reproduces the leading order terms in the corresponding TMD
factorization formalism in the regime $\Lambda_{\rm QCD}\ll
q_\perp\ll P_\perp$. This is a nontrivial result, especially for
the single spin asymmetry, because the calculations for the quark
Sivers function at this order have to take into account the
perturbative expansion of the gauge link at next-to-leading order
(${\cal O}(g^2)$).

\section{The $q_\perp$ moments of the SSA}

The single spin dependent differential cross section term in
Eq.~(\ref{e4}) can be further simplified by taking a
moment in $q_\perp$. Such a moment was also considered in \cite{mulders}.
We can rewrite
\begin{equation}
\epsilon^{\alpha\beta}S_\perp^\alpha
q_\perp^\beta=|S_\perp|\frac{|P_\perp|}{|q_\perp|}\sin\phi_b\sin\delta
\ ,
\end{equation}
where $\phi_b$ is the so-called bi-sector angle of the two jets:
$\phi_b=(\phi_1+\phi_2)/2$ with $\phi_1$ and $\phi_2$ the
azimuthal angles of the two jets relative to the polarization
vector $\vec{S}_\perp$. The angle $\delta=\pi-(\phi_2-\phi_1)$ measures
how far the two jets are away from the back-to-back configuration. All these
azimuthal angles are defined in a frame in which the polarized proton
is moving in the $+z$ direction.

Since $|q_\perp|\approx |P_\perp||\sin\delta|$, the
$q_\perp$-moment of the asymmetry is also related to the
$\sin\delta$-moment. One interesting moment is the following:
\begin{eqnarray}
\int d^2\vec{q}_\perp \frac{|P_\perp|}{M_P}\sin\delta
\frac{d^5\Delta \sigma(S_\perp)}
{dy_1dy_2dP_\perp^2d^2\vec{q}_\perp} =\sum_{ab}
\frac{-g}{M_P}x_aT_F^a(x_a,x_a)x_bf_b(x_b) H_{ab\to cd}^{\rm
Sivers}(P_\perp^2) \ ,
\end{eqnarray}
where $T_F(x,x)$ is the twist-3 matrix element of the quark-gluon
correlation function defined in Eq.~(\ref{TF}) \cite{qiusterman}.
It is also related to the $k_\perp$-moment of the TMD quark Sivers
function, see Eq.~(\ref{e34})
\cite{BoeMulPij03,{Ratcliffe:2007ye}}. An advantage of taking this
moment of the asymmetry is that the transverse momentum integrals
of the various factors in the factorization formula decouple from
each other, without further assumptions for the
$k_\perp$-dependence of the TMD distributions. In fact, the SSA in
the $q_\perp$ moment of the dijet momentum imbalance is
effectively a physical quantity with only a {\it single} large
observed scale, $P_\perp$. For such an observable, the
conventional collinear factorization is more appropriate. In this
case, all partons' transverse momentum dependence is integrated
into the usual collinear parton distributions. The hard factors we
obtained above for all the partonic channels agree with those
given in \cite{mulders}, although a very different approach was
adopted there. In \cite{mulders}, in order to investigate the
$q_\perp$-moment of the single spin asymmetry, the appropriate
gauge links were derived and expanded to first order in $g$. Since
both approaches have included the initial/final state interactions
in their formalisms, they should agree with each other at this
order.

\section{Conclusion}

In this paper, we have studied the asymmetric production of two
jets in hadronic collisions.  Using the collinear
factorization approach in perturbative QCD, we calculated both
the spin-averaged and the spin-dependent differential cross section for
the hadronic production of dijets with momentum imbalance
$q_\perp$, in a kinematic region where $P_\perp \gg q_\perp \gg
\Lambda_{\rm QCD}$. At the first nonvanishing order, the momentum
imbalance is generated by radiating a gluon with transverse
momentum equal to the imbalance. In the limit when the imbalance
$q_\perp$ is much less than the averaged jet momentum $P_\perp$,
we derived the leading-power contribution to the cross sections in
the $q_\perp/P_\perp$ expansion when the radiated gluon is nearly
parallel to one of the incoming hadrons. We found that the perturbatively
calculated leading contributions to both the spin-averaged and
the spin-dependent cross sections can be factorized into a partonic
hard part which is a function of $P_\perp$, and into
perturbatively generated TMD
parton distributions at ${\cal O}(g^2)$.  Our results derived in
the collinear factorization approach in the limit $q_\perp/P_\perp
\to 0$ reproduce the same factorized expressions in
Eqs.~(\ref{e37}),(\ref{e39}) that were derived as the leading order
terms of the generalized TMD factorization formulas in
Eqs.~(\ref{e4}),(\ref{e2}) when the observed $q_\perp$ is solely
due to the perturbatively generated TMD parton distribution of one
of the incoming hadrons.

The consistency between the results derived in the collinear
factorization approach and those derived by expanding the
generalized TMD factorization formulas in Eqs.~(\ref{e4}),(\ref{e2})
to the same order naturally leads us to ask if the generalized TMD
factorization formulas in Eqs.~(\ref{e4}),(\ref{e2}) are actually
valid for the hadronic dijet production to all orders in
perturbative QCD when $P_\perp \gg q_\perp \gtrsim \Lambda_{\rm
QCD}$, like the TMD factorization formulas for Drell-Yan and
SIDIS. {\it If this were true}, the
consistency that we demonstrated in this paper in the overlap
region where $P_\perp \gg q_\perp$ would effectively provide a unified
picture and treatment for the hadronic dijet production over the full
kinematic range of jet momenta: using the TMD factorization
formalism when the dijet momentum imbalance is small and the
collinear factorization approach when it is large.  This
is exactly what was achieved in Ref.~\cite{JiQiuVogYua06} for
the Drell-Yan and SIDIS processes.

However, as we mentioned at the beginning of Sec.~III, showing the
consistency between two factorized formalisms at the leading order
when $q_\perp/P_\perp \to 0$ is not sufficient to prove the TMD
factorization formula over its full kinematic region.  When we
started with the collinear factorization approach to calculate the
momentum imbalance of the dijet cross section, we already took as
a fact that the long-distance interaction between the hadrons/jets
at their mass scales, ${\cal O}(\Lambda_{\rm QCD})$, are
factorized for the given $q_\perp\sim P_\perp$. That is, our
results derived in the limit $q_\perp/P_\perp \to 0$ are valid
when $q_\perp \gg \Lambda_{\rm QCD}$; but, our derivation does not
address the issues concerning the factorization of the
interactions between the hadrons/jets when $q_\perp \to
\Lambda_{\rm QCD}$ or the mass scale of the hadrons/jets involved.
The consistency that we demonstrated here is certainly a
necessary condition that needs to be satisfied should the TMD factorization
formulas be valid.

For the TMD factorization to be true, we need to establish two key
facts: the partonic hard part at ${\cal O}(P_\perp)$ must be
insensitive to any physics at the scale $q_\perp\ll P_\perp$, and
the TMD parton distributions at the scale $q_\perp$ should be
process independent and universal, or at least, ``quasi''-universal
up to a sign change \cite{BroHwaSch02,Col02}. In this paper, we
also calculated in Sec.~II the partonic hard parts at ${\cal
O}(P_\perp)$ by using the Brodsky-Hwang-Schmidt model with the
proper color factor for the interaction vertices.  We derived the
partonic hard parts from the lowest order scattering diagrams and
showed that the dynamics of partonic scattering at scale ${\cal
O}(P_\perp)$ is independent of the model of nucleon that we used
in our derivation.  Since the SSA can be generated by either
initial- or final-state interactions, the color flow of the
partonic scattering in the dijet momentum imbalance is very
different from that of Drell-Yan and SIDIS. Since all poles from
the initial- and final-state one gluon interactions contribute to
the SSA, and the color factor for each individual scattering diagram is
insensitive to the long-distance details of the nucleon because
there is only one unique color structure for the active
quark-gluon combination at this order, we were able to factor the
process-dependent as well as the diagram-specific color factors along with
the partonic hard parts, and leave the physics at scale ${\cal
O}(q_\perp)$ in the TMD parton distributions defined in SIDIS. By
doing that, we found that the resulting full partonic hard parts
are the same as those derived in
the collinear factorization approach. However, for a full TMD
factorization, we will have to show that the poles from multiple
gluon interactions can be exponentiated into the gauge link that
defines the TMD parton distributions, while leaving the same partonic
hard parts with the same color factors.  Due to the mixture of
initial- and final-state interactions and the complications of color
flows, this task is very non-trivial and is beyond the scope of
this paper.

To better understand the dijet momentum imbalance in hadronic
collisions, a number of extensions can be performed based on our
results. First, in this paper, we have only studied the
contribution of one gluon radiation collinear to one of the
incident hadrons. This should be extended to other kinematic
regions of the gluon.  For example, when the gluon is radiated
parallel to the final state jets or hadrons, it is important to
see how the corresponding contribution can be accommodated by the
jet definition, or be factorized into the fragmentation functions
for the hadron, especially in the spin-dependent case. Another
important kinematic regime is when the radiated gluon becomes
soft. It is very important to see if the resulting contribution
can be factorized into a soft factor, as predicted by the TMD
factorization formalism. In addition, in this paper we have only
discussed the SSA for dijet correlations, which has Sivers-type
contributions. For the related di-hadron correlations, one expects
also the Collins mechanism to be important, and it will be
interesting to explore it and the factorization properties of the
corresponding spin-dependent cross section. The method and
procedure used in this paper should provide useful guidance in all
these related studies.

We finally note that if the TMD factorization indeed fails, a careful
comparison between our results and experimental data as the
momentum imbalance $q_\perp/P_\perp$ decreases could furnish an ideal test
of the breaking of factorization and provide new opportunities to
explore QCD dynamics.

\section*{Acknowledgments}
We are grateful to Cedran Bomhof, John Collins, Andreas Metz, and
Piet Mulders for useful discussions. We thank Xiangdong Ji for his
general remarks and his reminding us the Wigner-Eckart theorem for
the color-factor decomposition. We especially thank George Sterman
for continued discussions on the factorization issues in hadronic
reactions. J.Q. is supported in part by the U. S. Department of
Energy under grant No. DE-FG02-87ER-40371. W.V. and F.Y. are
finally grateful to RIKEN, Brookhaven National Laboratory and the
U.S. Department of Energy (contract number DE-AC02-98CH10886) for
providing the facilities essential for the completion of their
work. J.Q. thanks the high energy theory group at Argonne National
Laboratory for its hospitality during the writing of this work.

\appendix
\section{}
In this appendix, we list the hard factors for the
dijet-correlation at hadron colliders, for the unpolarized and the
single-transverse spin dependent cases. For unpolarized
scattering, the hard factors are well-known in the literature.
They may also be obtained from the results shown in Tables
I-III, using Eq.~(\ref{e29}). For completeness, we list them here:
\begin{eqnarray}
H_{qq'\to qq'}^{uu}(\hat s,\hat t,\hat u)&=&H_{q\bar q'\to q\bar
q'}^{uu}=\frac{\alpha_s^2\pi}{\hat
s^2}\frac{N_c^2-1}{4N_c^2}\frac{2(\hat s^2+\hat u^2)}{\hat
t^2}\ ,\nonumber\\
H_{q\bar q\to q'\bar q'}^{uu}(\hat s,\hat t,\hat
u)&=&\frac{\alpha_s^2\pi}{\hat
s^2}\frac{N_c^2-1}{4N_c^2}\frac{2(\hat t^2+\hat
u^2)}{\hat s^2} \ ,\nonumber\\
H_{qq\to qq}^{uu}(\hat s,\hat t,\hat
u)&=&\frac{\alpha_s^2\pi}{\hat
s^2}\left\{\frac{N_c^2-1}{4N_c^2}\left[\frac{2(\hat s^2+\hat
u^2)}{\hat t^2}+\frac{2(\hat s^2+\hat t^2)}{\hat
u^2}\right]-\frac{N_c^2-1}{4N_c^3}\frac{4\hat s^2}{\hat t\hat
u}\right\} \
,\nonumber\\
H_{q\bar q\to q\bar q}^{uu}(\hat s,\hat t,\hat
u)&=&\frac{\alpha_s^2\pi}{\hat
s^2}\left\{\frac{N_c^2-1}{4N_c^2}\left[\frac{2(\hat s^2+\hat
u^2)}{\hat t^2}+\frac{2(\hat u^2+\hat t^2)}{\hat
s^2}\right]-\frac{N_c^2-1}{4N_c^3}\frac{4\hat u^2}{\hat s\hat
t}\right\} \ ,\nonumber\\
H_{qg\to qg}^{uu}(\hat s,\hat t,\hat
u)&=&\frac{\alpha_s^2\pi}{\hat s^2}\left\{\frac{1}{2}\frac{2(\hat
s^2+\hat u^2)}{\hat t^2}-\frac{C_F}{2N_c}\frac{2(\hat s^2+\hat
u^2)}{\hat
s\hat t}\right\} \ ,\nonumber\\
H_{q\bar q\to gg}^{uu}(\hat s,\hat t,\hat
u)&=&\frac{\alpha_s^2\pi}{\hat
s^2}\left\{\frac{C_F^2}{N_c}\frac{2(\hat t^2+\hat u^2)}{\hat t\hat
u}-C_F\frac{2(\hat t^2+\hat u^2)}{\hat s^2}\right\} \ ,\nonumber
\end{eqnarray}
\begin{eqnarray}
H_{qg\to q\gamma}^{uu}(\hat s,\hat t,\hat
u)&=&\frac{\alpha_s\alpha_e e_q^2\pi}{\hat
s^2}\frac{1}{2N_c}\frac{2(\hat s^2+\hat u^2)}{-\hat s\hat
u} \ ,\nonumber\\
H_{q\bar q\to g\gamma}^{uu}(\hat s,\hat t,\hat
u)&=&\frac{\alpha_s\alpha_e e_q^2\pi}{\hat
s^2}\frac{C_F}{N_c}\frac{2(\hat t^2+\hat u^2)}{\hat t\hat u} \ .
\end{eqnarray}
For the single-transverse spin
dependent case, using Eq.(\ref{e30}), we obtain
\begin{eqnarray}
\label{hsivers} H_{qq'\to qq'}^{\rm Sivers}(\hat s,\hat t,\hat
u)&=&\frac{\alpha_s^2\pi}{\hat
s^2}\frac{N_c^2-5}{4N_c^2}\frac{2(\hat s^2+\hat u^2)}{\hat
t^2}\ ,\nonumber\\
H_{q\bar q'\to q\bar q'}^{\rm Sivers}(\hat s,\hat t,\hat
u)&=&\frac{\alpha_s^2\pi}{\hat
s^2}\left(-\frac{N_c^2-3}{4N_c^2}\right)\frac{2(\hat s^2+\hat
u^2)}{\hat t^2}\ , \nonumber\\
H_{q\bar q\to q'\bar q'}^{\rm Sivers}(\hat s,\hat t,\hat
u)&=&\frac{\alpha_s^2\pi}{\hat
s^2}\frac{N_c^2+1}{4N_c^2}\frac{2(\hat t^2+\hat
u^2)}{\hat s^2} \ ,\nonumber\\
H_{qq\to qq}^{\rm Sivers}(\hat s,\hat t,\hat
u)&=&\frac{\alpha_s^2\pi}{\hat
s^2}\left\{\frac{N_c^2-5}{4N_c^2}\left[\frac{2(\hat s^2+\hat
u^2)}{\hat t^2}+\frac{2(\hat s^2+\hat t^2)}{\hat
u^2}\right]+\frac{N_c^2+3}{4N_c^3}\frac{4\hat s^2}{\hat t\hat
u}\right\} \
,\nonumber\\
H_{q\bar q\to q\bar q}^{\rm Sivers}(\hat s,\hat t,\hat
u)&=&\frac{\alpha_s^2\pi}{\hat
s^2}\left\{-\frac{N_c^2-3}{4N_c^2}\frac{2(\hat s^2+\hat u^2)}{\hat
t^2}+\frac{N_c^2+1}{4N_c^2}\frac{2(\hat u^2+\hat t^2)}{\hat
s^2}-\frac{N_c^2+1}{4N_c^3}\frac{4\hat u^2}{\hat s\hat
t}\right\} \ ,\nonumber\\
H_{qg\to qg}^{\rm Sivers}(\hat s,\hat t,\hat
u)&=&\frac{\alpha_s^2\pi}{\hat
s^2}\left\{-\frac{N_c^2}{4(N_c^2-1)}\frac{2(\hat s^2+\hat
u^2)}{\hat t^2}\left[\frac{\hat s}{\hat u}-\frac{\hat u}{\hat
s}\right]-\frac{1}{2(N_c^2-1)}\frac{2(\hat s^2+\hat u^2)}{\hat
t^2}\right.\nonumber\\
&&~~~~~~~\left.-\frac{1}{4N_c^2(N_c^2-1)}\frac{2(\hat s^2+\hat
u^2)}{\hat
s\hat u}\right\} \ ,\nonumber\\
H_{q\bar q\to gg}^{\rm Sivers}(\hat s,\hat t,\hat
u)&=&\frac{\alpha_s^2\pi}{\hat
s^2}\left\{-\frac{1}{2N_c}\frac{2(\hat t^2+\hat u^2)}{\hat
s^2}+\frac{N_c}{4}\frac{2(\hat t^2+\hat u^2)}{\hat
s^2}\left[\frac{\hat t}{\hat u}+\frac{\hat u}{\hat t}\right]\right.\nonumber\\
&&~~~~~~~\left.-\frac{2N_c^2+1}{4N_c^3}\frac{2(\hat t^2+\hat
u^2)}{\hat t\hat u}\right\} \ ,\nonumber\\
H_{qg\to q\gamma}^{\rm Sivers}(\hat s,\hat t,\hat
u)&=&\frac{\alpha_s\alpha_e e_q^2\pi}{\hat
s^2}\left(-\frac{N_c^2+1}{2N_c(N_c^2-1)}\right)\frac{2(\hat
s^2+\hat u^2)}{-\hat s\hat u} \ ,\nonumber\\
H_{q\bar q\to g\gamma}^{\rm Sivers}(\hat s,\hat t,\hat
u)&=&\frac{\alpha_s\alpha_e e_q^2\pi}{\hat
s^2}\left(\frac{N_c^2+1}{2N_c^2}\right)\frac{2(\hat t^2+\hat
u^2)}{\hat t\hat u} \ .
\end{eqnarray}
The above results are for $t$-channel scattering. In
dijet-production, both $t$ and $u$ channel processes will
contribute. The hard factors for the $u$-channel scattering can be
obtained from the results above by exchanging $\hat t\leftrightarrow
\hat u$.


\end{document}